\documentclass{article}  

\usepackage{booktabs}
\usepackage{amsmath}
\usepackage{authblk}
\usepackage{graphicx}
\usepackage{abstract}
\usepackage{amssymb}
\usepackage{subcaption}
\usepackage{fancyhdr}
\usepackage{hyperref}
\usepackage{color}
\usepackage[round]{natbib}
\usepackage{lineno}

\title{A proof of concept for scale-adaptive parameterizations: the case of the Lorenz '96 model}

\author[1,2]{Gabriele Vissio}
\author[2,3,4]{Valerio Lucarini}

\affil[1]{International Max Planck Research School on Earth System Modelling, Hamburg, Germany}
\affil[2]{CEN, Meteorological Institute, University of Hamburg, Hamburg, Germany}
\affil[3]{Department of Mathematics and Statistics, University of Reading, Reading, UK}
\affil[4]{Walker Institute for Climate System Research, University of Reading, Reading, UK}
\begin{document}
\maketitle
\begin{onecolabstract}
Constructing efficient and accurate parameterizations of sub-grid scale processes is a central area of interest in the numerical modelling of geophysical fluids. Using a modified version of the two-level Lorenz '96 model, we present here a proof of concept of a scale-adaptive parameterization constructed using statistical mechanical arguments. By a suitable use of the Ruelle response theory and of the Mori-Zwanzig projection method, it is possible to derive explicitly a parameterization for the fast variables that translates into deterministic, stochastic and non-markovian contributions to the equations of motion of the variables of interest. {\color{black}We show that our approach is computationally parsimonious,} has great flexibility, as it is explicitly scale-adaptive, and we prove that it is competitive compared to empirical ad-hoc approaches. While the parameterization proposed here is universal and can be easily analytically adapted to changes in the parameters' values by a simple rescaling procedure, the parameterization constructed with the ad-hoc approach needs to be recomputed each time {\color{black}the parameters of the systems are changed.} The price of the higher flexibility of the method proposed here is having a lower accuracy in each individual case.
\end{onecolabstract}

{\bf Keywords:} Parameterization; Multiscale systems; Stochastic Dynamics; Memory; Noise; Response theory; Mori-Zwanzig theory; Chaos; Scale-adaptivity; Prediction

\section{Introduction}
The climate is a forced and dissipative system featuring variability on a vast range of spatial and temporal scales. This results essentially from the fact that a) the climate system is composed by subdomains having different characteristic time scales; and b) the dynamics inside each subdomain and the couplings between them are strongly nonlinear. As a result, even the most sophisticated and computationally expensive numerical climate models are far from being able to represent explicitly just a relatively small fraction of the whole dynamical range {\color{black}of the geophysical fluids} \citep{Ghil1987,Peixoto1993,Lucarini2014}. Therefore, it is crucial to develop approximate - yet accurate and efficient - dynamical/statistical representations - the so-called parameterizations - of the effects of unresolved scales on the scales the model is able to explicitly describe \citep{Palmer2008,Franzke2015,Berner2016}. Lacking proper parameterizations reduces substantially model's skills in terms of short-to-medium range weather prediction, and on climatic time scales, in terms of average properties, variability, and climate response to forcings \cite[see e.g.][]{Holton2004,McGuffie2005,Palmer2006,Plant2016}.

A fundamental problem in the construction of parameterizations is that they are typically tuned for being accurate for a specific configuration of a model in terms of numerical resolution, and the operation of re-tuning needed when a new model version at higher resolution is available can be extremely tedious and costly. The need of achieving scale-adaptive parameterizations has been recently emphasized in the scientific literature, see e.g. \cite{Arakawa2011,Park2014,Sakradzija2016}. Additionally, parameterizations are typically tested against {\color{black}specific observables of interest and tuned in order to better represent those observables}, but it is not always clear whether optimizing the skill for such observables might come at the price of reducing the skill on other climatic properties that might prove crucial for, e.g., modulating the climatic response to forcings.

A somewhat peculiar point of view is provided by the so-called superparameterizations, mostly used to represent convection \citep{Majda2007,Li2012}. The idea is to have lower dimensionality (and so computationally much cheaper) models run in parallel with the main code for resolving at high resolution the dynamics inside each atmospheric column.

For long time, parameterizations were aimed at describing the mean effects of small, fast scales on slower ones, but recently it has become apparent that it is crucial to widen their formulation in order be able to include in some way the effect of fluctuations. The pursuit of stochastic parameterizations for weather and climate models has then become an extremely active area of research, see e.g. the recent contributions by \cite{Palmer2008,Franzke2015,Berner2016} and the now classical collection of results in \cite{Imkeller2001}. The construction of stochastic parameterizations in geophysical fluid dynamical models is also usually approached using a pragmatic method: one tries to construct empirical functions able to represent well the effect of mean state and of the fluctuations of the unresolved variables, see e.g. the illustrative examples of \cite{Orrell2003} and \cite{Wilks2005}.

Mathematical arguments do indeed support the idea of going towards stochastic parameterizations. The homogenization method shows that the effect of the fast scales on the slow scales can be represented, in the limit of infinite time scale separation, as the sum of two extra terms in the equation of motions of the slow variables, precisely a deterministic (mean field) and a white-noise stochastic (fluctuations) component \citep{Pavliotis2008}. {\color{black}This point of view has led to an important set of results by Majda and collaborators on the possibility of constructing explicitly reduced order models for geophysical fluid dynamical systems (see, e.g., \cite{Majda1999,Majda2001,Majda2003,Franzke2005}).

A different point of view focuses, instead, on constructing effective dynamics comprising deterministic as well as stochastic terms purely from data. The idea proposed by \cite{Kravtsov2005} has been to extend the multilevel linear regressive method, which is suitable for linear problems, to the nonlinear case, allowing for dealing with the possibility of representing quadratic nonlinearities in the evolution equations, which are in fact typical of (geophysical) fluid dynamical processes. The method allows for constructing an optimal representation of the deterministic, linear and nonlinear, dynamics as well as of the stochastic forcing, so that its correlation properties are suitably recovered without making any assumption on the existence of time scale separation between resolved and neglected variables. }

The Mori-Zwanzig theory \citep{Zwanzig1960,Zwanzig1961,Mori1974} provides an - unfortunately implicit - expression for the effect of the small, fast scales on the scales of interest. One finds that, once the hypothesis of infinite time scale separation is relaxed, the parameterization requires in fact three terms, a deterministic correction, a stochastic term, and a memory, non-markovian term. In the limit of infinite time scale separation, the memory term drops off and the stochastic terms becomes indeed white noise, in agreement with what predicted by the homogenization theory.

Recently \cite{Chekroun2015a,Chekroun2015b} have provided a comprehensive treatment of these issues that combine mathematical rigor and physical intuition. In a few recent papers, \cite{Wouters2012,Wouters2013,Wouters2016} have provided explicit formulas for constructing parameterizations able to incorporate the deterministic, stochastic, and non-markovian components. The formulas have been obtained independently using two rather different approaches, namely a second order expansion of the Mori-Zwanzig projection operator, and a reworking of the \cite{Ruelle1998,Ruelle2009} response theory, which allows under suitable conditions to compute the change in the expectation value of any smooth observable of a system resulting from perturbations of the dynamics in terms of the statistical properties of the unperturbed flow. 

The idea followed by Wouters and Lucarini has been to treat the coupling between the slow and fast variables as the weak forcing added on top of the uncoupled dynamics, and then evaluate the impact of the forcings on the statistical properties of a generic observable of the slow variables. Finally, the last step has been to retro-engineer explicit formulas for terms that, added to the uncoupled dynamics of the slow variables, provide up to second order to the same results as the actual coupling.

It is important to note that since explicit formulas are provided, one can indeed construct the parameterizations \textit{ab-initio}, and not empirically. Additionally, the parameterization is automatically optimized for all possible observables of the system. Such an approach seems especially promising in all systems, as in the extremely relevant case of the climate, where there is no \textit{spectral gap} in the scales of motions that justifies the assumption of infinite time scale separation between fast and slow scales. It seems then in general relevant to be able to retain and check the relevance of the memory term and to construct a suitable model for the stochastic forcing, going beyond the approximation of using white noise or simple empirical autoregressive processes. 

Note that the approach discussed here is not \textit{per se} constructed to deal with multiscale systems only. In fact, the explicit expressions for the terms responsible for the parameterization are constructed by performing an asymptotic expansion controlled by a parameter determining the degree of coupling between the set of variables of interest and those we want to parameterize. Clearly, if such a condition is satisfied, we can apply our method also to multiscale systems, as done here. Recently, a parameterization constructed according to such a statistical mechanical point of view has been tested successfully in a simple low dimensional model \citep{Wouters2016a} and in a more complex yet simple coupled model \citep{Demaeyer2015}. In this paper, we want to stress another quality of this approach, namely the possibility of having automatically scale adaptive formulations of the parameterization. 

We choose as benchmark system to work with (a modified version of) the Lorenz '96 model \citep{Lorenz1996}, which provides a prototypical yet convincing representation of a two-scale system where large scale, synoptic variables are coupled to small scale, convective variables. The Lorenz '96 model has quickly become the test-bed for evaluating new methods of data assimilation \citep{Trevisan2004,Trevisan2010} and is receiving a lot of attention also in the community of statistical physics \citep{Abramov2008,Hallerberg2010,Lucarini2011,Gallavotti2014}. More importantly for our specific case, the Lorenz '96 model has been used in the papers of \cite{Orrell2003} and \cite{Wilks2005} to construct explicit models of stochastic parameterization, so we have previous results to compare to. {\color{black}We wish to stress that data driven closure models have been recently extended in order to be able to deal with the unavoidable memory effects due to presence of neglected, hidden variables \citep{Kondrashov2017}. One can see the approach proposed here as the top-down counterpart of the bottom-up approach provided by the data-driven methods. }

The paper is structured as follows. Section 2 provides the main ingredients of the method for constructing general parameterizations introduced by \cite{Wouters2012,Wouters2013,Wouters2016}. Section 3 describes the Lorenz '96 system and highlights the modifications we have applied in the present work, most importantly the introduction of a forcing acting also on the fast variables. Section 4 is dedicated to describing the performance of the parameterization, discuss its scale-adaptive properties, and compare its performance with previous results. Section 5 concludes the paper with the discussion of the results and the perspective for future research in this area. In Appendix A we present some new ideas building on Wouters and Lucarini 2012, 2013, 2016 able to extend the range of applicability of the theory.

\section{Wouters-Lucarini's parameterization}
In this paper we test the effectiveness of the methodology introduced by \cite{Wouters2012,Wouters2013,Wouters2016} for constructing parameterizations for dynamical systems of the form:
\begin{align}
  \frac{dX}{dt}&=F_X(X)+\epsilon\Psi_X(X,Y),\label{eq:perturbedX}\\
  \frac{dY}{dt}&=F_Y(Y)+\epsilon\Psi_Y(X,Y),\label{eq:perturbedY}
\end{align}
where the $X$ variables correspond to the dynamics of interest and the $Y$ variables correspond to the dynamics we want to parameterize. The $F$ vector field on the right hand side of Eqs. \eqref{eq:perturbedX}-\eqref{eq:perturbedY} corresponds to the uncoupled dynamics of the $X$ and $Y$ variables respectively, while the $\Psi$ field describes the coupling, with $\epsilon$ being a bookkeeping variable describing the coupling strength. Note that Eqs. \eqref{eq:perturbedX}-\eqref{eq:perturbedY} do not describe, in general, a multiscale dynamical system, where the $X$ (slow) and the $Y$ (fast) variables are essentially characterized by different scales of motion. Nonetheless, we can bring it to the standard form elucidating multiscale behaviour by considering the following form for Eqs. \eqref{eq:perturbedX}-\eqref{eq:perturbedY}:
\begin{equation} \label{newperturbedX}
\frac{dX}{dt}=F_X(X)+\epsilon\Psi_X(X,Y)
\end{equation}
\begin{equation} \label{newperturbedY}
\frac{dY}{dt}=\gamma\tilde{F}_Y(Y)+\epsilon{\Psi}_Y(X,Y)
\end{equation}
where $\gamma\gg1$ and $F_Y(Y)=\gamma \tilde{F}_Y(Y)$. As clear from the later discussion, it is not important in our case to include the factor $\gamma$ also for the coupling term affecting the $Y$ variables in Eq. \eqref{newperturbedY}, because we are eminently interested in separating the time scales of the decoupled ($\epsilon=0$) X- and Y-systems. Following the discussion presented in the introduction, the goal is to find an approximate equation of the form
\begin{equation} \label{eq:Xpluschi}
  \frac{dX}{dt}=F_X(X)+\chi\{X\}
\end{equation}
able to provide a good approximation of the statistical properties of the $X$ variables, where $\chi\{X\}$ can in general correspond to an integro-differential contribution with also a stochastic component. It seems relevant aiming at being able to specify in advance the accuracy of the approximation in terms of the properties of the coupling and, in particular, of the coupling strength $\epsilon$. Clearly, if $\epsilon=0$, we have that $\chi\{X\}=0$ provides a (trivial) solution to our problem. The approach can be seamlessly followed also in the presence of a functional form for the equations where the parameter $\gamma$ explicitly controls the scale separation between the $X$ and $Y$ variables. Note that \cite{Abramov2016} has recently introduced an extension of the homogenization method able to deal with a problem formulated as in Eqs. \eqref{eq:perturbedX}-\eqref{eq:perturbedY}.

\subsection{The method}

The basic idea is to consider the dynamical system \eqref{eq:perturbedX}-\eqref{eq:perturbedY} as resulting from an $\epsilon-$perturbation of the following dynamical system:
\begin{equation} \label{eq:unperturbedX}
\frac{dX}{dt}=F_X(X),
\end{equation}
\begin{equation} \label{eq:unperturbedY}
\frac{dY}{dt}=F_Y(Y),
\end{equation}
where the coupling plays the role of the perturbation. We now focus on the $X$ variables by considering a general observable $A=A(X)$, i.e. a smooth function of the $X$ variables only. Making suitable hypotheses on the mathematical properties of the unperturbed system and taking advantage of the  \cite{Ruelle1998,Ruelle2009} response theory, Wouters and Lucarini have been able to find a useful expression for the expectation value $\rho_\epsilon(A)$ of the observable $A$ taken according to the invariant measure $\rho_\epsilon(dXdY)$ of the coupled dynamical system \eqref{eq:perturbedX}-\eqref{eq:perturbedY}:
\begin{equation}
  \rho_\epsilon(A)=\int\rho_\epsilon(dXdY)A(X) .
\end{equation}
In what follows, we assume that all invariant measures considered are of the Sinai-Ruelle-Bowen kind \citep{Eckmann1985,Young2002}. This assumption can be physically motivated by taking into account the chaotic hypothesis (e.g. \cite{Gallavotti2014a}). We can also introduce the projected measure 
\begin{equation}
\rho^*_\epsilon(dX)=\int_Y \rho_\epsilon(dX dY)
\end{equation}
where the inetgration is performed on the $Y$ variables only, such that $\rho_\epsilon(A)=\rho^*_\epsilon(A)$. Using ergodicity, we also have:
\begin{equation} \label{ensambleaverage}
\rho^*_\epsilon(A)=\lim_{T\to\infty} \frac{1}{T}\int_0^Td\tau A(x(t)) ,
\end{equation}
where $x(t)=\widetilde f^t{x_0}$, with $\widetilde f^t$ defining the flow determined by the dynamical system \eqref{eq:perturbedX}-\eqref{eq:perturbedY}. It is possible to find a perturbative expansion of the expectation value of $A$ taken according to the invariant measure of the coupled system. One can in fact write 
\begin{equation}
  \rho^*_\epsilon(A)=\rho_{0,X}(A)+\epsilon \delta_\Psi^{(1)}\rho(A)+\epsilon^2 \delta_{\Psi,\Psi}^{(2)}\rho(A)+\mathcal{O}(\epsilon^3) ,
\end{equation}
where the first term $\rho_{0,X}(A)$ is the expectation value of $A$ taken according to the invariant measure of the $X$-component of the unperturbed system \eqref{eq:unperturbedX}:
\begin{equation}
  \rho_{0,X}(A)=\int\rho_{0,X}(dX)=\lim_{T\to\infty} \frac{1}{T}\int_0^Td\tau A(f^\tau(x_0)) ,
\end{equation}
where we have again used ergodicity and defined $f^t$ as the flow of the $X$ variables part of the dynamical system \eqref{eq:unperturbedX}. The second term $\epsilon \delta_\Psi^{(1)}\rho(A)$ and the third term $\epsilon^2 \delta_{\Psi,\Psi}^{(2)}\rho(A)$ correspond to the first and second order corrections, and can be also expressed as expectation values on $\rho_{0,X}(dX)$ of explicitly determined observables, which are constructed non-trivially from A and the vector field $\Psi$. All the terms can be computed from the statistical properties of the uncoupled dynamics of the Y variables given in Eq. \ref{eq:unperturbedY}. The explicit expressions can be found in \cite{Wouters2012}.

While the previous result allows for computing the impact of the coupling on the statistics of any given $A$ observable, it is not useful \textit{per se} for constructing a parameterization. Nonetheless, it is possible to retro-engineer an educated guess for the term $\chi\{X\}$ introduced in Eq. \eqref{eq:Xpluschi}, such that \textit{up to second order in $\epsilon$} the expectation value of $A$ according to the invariant measure $\rho'_\epsilon(dX)$ of the system:
\begin{equation} \label{eq:wl_par}
  \frac{dX}{dt}=F_X(X)+\epsilon D(X)+\epsilon S\{X\}+\epsilon^2 M\{X\}
\end{equation}
is the same as the expectation value of $A$ according to $\rho_\epsilon$, or, more explicitly:
\begin{equation} \label{eq:rhoOthird}
  \rho_\epsilon(A)=\rho'_\epsilon(A)+O(\epsilon^3).
\end{equation}  

Therefore, Eq. \eqref{eq:rhoOthird} provides a useful basis for defining a parameterization where we are able to control the error on the statistics of the surrogate dynamics with respect to the full dynamics as a function of $\epsilon$, and where this applies \textit{for all possible observables} $A$. 

The three perturbation vector fields $D$, $S$ and $M$ correspond to, respectively, a mean field term, a stochastic forcing and a non-markovian memory term. Note that the stochastic term has a second order effect on the measure even if its intensity is proportional to $\epsilon$; see \cite{Lucarini2012}. As shown in
\cite{Wouters2012,Wouters2013,Wouters2016}, the explicit expression for these three terms can be obtained also by performing a second order expansion of the Mori-Zwanzig projector operator, which constructs the effective projected dynamics for the $X$ variables only. This suggests that the proposed parameterization might have skill also in terms of prediction (in the sense of weather forecast); we will test this elsewhere. In what follows, we refer to this approach as the the W-L parameterization.

The explicit expressions for the three terms providing the parameterization shown in  Eq. \eqref{eq:wl_par} are given below in Eqs. \eqref{eq:original_D}, \eqref{eq:Smult} and \eqref{eq:original_h}. Therefore, once we derive $D$, $S$, and $M$, we can use them to construct parameterizations for all values of $\epsilon$ within the radius of convergence of the expansion. Additionally, if the coupled model given in Eqs. \eqref{eq:perturbedX}-\eqref{eq:perturbedY} is multiscale, this approach allows for constructing parameterizations integrating the single scale equation \eqref{eq:unperturbedY}. This can significantly ease the computational burden of our problem.

\subsubsection{Deterministic, stochastic, and non-markovian terms}
 
We assume that the coupling terms $\Psi_X(X,Y)$ and $\Psi_Y(X,Y)$ are separable in the $X$ and $Y$ variables, so that we can write $\Psi_X(X,Y)=\Psi_{X,1}(X)\Psi_{X,2}(Y)$ and $\Psi_Y(X,Y)=\Psi_{Y,1}(X)\Psi_{Y,2}(Y)$. As explained in \cite{Wouters2012,Wouters2013,Wouters2016}, such an assumption does not really impact the generality of our results.

$D(X)$ is a deterministic term that accounts for the average impact that the coupling has on the $X$ variables and it is given by:

\begin{equation} \label{eq:original_D}
  D(X)=\Psi_{X,1}(X)\rho_{0,Y}(\Psi_{X,2}(Y)).
\end{equation}

The second order contribution is composed of two parts. $S\{X\}$ represents a stochastic forcing due to the temporal correlation of the fluctuations of the forcing exerted by the Y-variables onto the $X$ variables. 
We can write
\begin{equation} \label{eq:Smult}
  S\{X\}=\Psi_{X,1}(X)\sigma(t) ,
\end{equation}
where $\sigma(t)$ is a stochastic term 
and is constructed in such a way to reproduce the lagged correlation of the fluctuations of the forcing. The statistical properties of the noise $\sigma(t)$ can be expressed as:
\begin{align}
  R(t)&=\langle\sigma(t),\sigma(0)\rangle\nonumber\\
  &=\rho_{0,Y}\left((\Psi_{X,2}(Y)-\rho_{0,Y}(\Psi_{X,2}(Y)))(\Psi_{X,2}(f^{t}(Y))-\rho_{0,Y}(\Psi_{X,2}(Y)))\right),\\
  \langle\sigma(t)\rangle&=0.
\end{align}
where the brackets indicate the expectation value of the stochastic process and $R(t)$ is the lagged correlation of the (stationary) noise. 

Finally, $M\{X\}$ is a memory term that describes the effects of the history of the $X$ variables on their present value through the influence of the $Y$ variables. This term is essential for capturing the effect of the hidden (Y) variables on the (X) variables of interest, as clarified by \cite{Chekroun2015a,Chekroun2015b}. It is expressed as:
\begin{equation}
  M\{X\}=\int_0^\infty h(\tau,X(t-\tau))d\tau,
\end{equation}
where the integral kernel is given by:
\begin{equation} \label{eq:original_h}
  {\color{black}h(\tau,\tilde{X})=\Psi_{Y,1}(\tilde{X})\Psi_{X,1}(f^{\tau}(\tilde{X}))\rho_{0,Y}(\Psi_{Y,2}(Y)\partial_Y\Psi_{X,2}(f^{\tau}(Y))).}
\end{equation}
Such an average resembles a cross-correlation between the actual state of the two fields $X,Y$ and the deviation of the trajectory of the same fields evolved at $t=\tau$.
\newline A remarkable property of this parameterization is its universality, as shown by Eq. \eqref{eq:original_D} through \eqref{eq:original_h}, because we have explicit formulas for computing the three factors $D$, $S$ and $M$ for any given expression of the coupling terms or of the uncoupled dynamics. Another positive aspect of these equations is the scale adaptivity of the parameterization terms, as we are going to show in next sections.

\subsection{Independent coupling case}

The special case where the two coupling terms are independent from the variable they are affecting, namely $\Psi_X(X,Y)=\Psi_X(Y)$ and $\Psi_Y(X,Y)=\Psi_Y(X)$, 
is particularly important for the scopes of this paper. The three terms discussed above take the following simpler form:
\begin{equation} \label{eq:Dterm}
  D(X)=\rho_{0,Y}(\Psi_X(Y)),
\end{equation}
\begin{equation}  \label{eq:m2}
  S\{X\}=\sigma(t),
\end{equation}
where
{\color{black}
\begin{equation} \label{eq:fluctuationterm}
\begin{split}
  R(t)=\langle\sigma(0),\sigma(t)\rangle&=\rho_{0,Y}((\Psi_X(Y)-D)(\Psi_X(f^{t}(Y))-D)),\\
  \langle\sigma(t)\rangle&=0,
\end{split}
\end{equation}}
and
\begin{equation} \label{eq:m3}
  M\{X\}=\int_0^\infty h(t_2,X(t-t_2))dt_2,
\end{equation}
where
{\color{black}
\begin{equation} \label{eq:new_h}
  h(t_2,\tilde{X})=\Psi_Y(\tilde{X})\rho_{0,Y}(\partial_Y\Psi_X(f^{t_2}(Y))).
\end{equation}}
In this special case, the stochastic contribution reduces to a simple additive noise term - compare Eqs. \eqref{eq:Smult} and \eqref{eq:m2} - while the evaluation of the memory kernel $h$ is significantly easier as a simpler expression appears in the ensemble average - compare Eqs. \eqref{eq:original_h} and \eqref{eq:new_h}.

We can now prove the scale adaptivity of the method adopted here as follows. We then consider the case where the equations of motions can be written as in \eqref{eq:unperturbedX}-\eqref{eq:unperturbedY}. We note that the expectation values are computed according to the invariant measure of the uncoupled equation $\frac{dY}{dt}=\gamma\tilde{F}_Y(Y)$, which can be rewritten as 
\begin{equation} \label{universa}
\frac{dY}{d\tau}=\tilde{F}_Y(Y)
\end{equation} 
where $\tau=\gamma t$. 

We clearly have that the constant $D$ in Eq. \eqref{eq:Dterm} is not affected by the choice of the time scale. Instead, the correlation function in Eq. \eqref{eq:fluctuationterm} and the memory kernel in Eq. \eqref{eq:new_h} are affected by the rescaling in the time and only the rescaled time $\tau$ will appear in their arguments. By substituting $\tau=ct$ one then obtains the actual parameterization for every choice of $\gamma$. In particular, large values of $\gamma$ will lead to a compression of the time axis for the correlation function and memory kernel, as seen below in the specific case investigated in this study.

\section{The Lorenz '96 Model}

It is crucial to test the methodology outlined in the previous section on a concrete numerical model having some practical and conceptual relevance. We recommend the reader to check the recent contributions by \cite{Wouters2016a} and \cite{Demaeyer2015}. The study presented here is constructed in such a way that we systematically explore how the performance of the parameterization changes when we alter both the intensity of the coupling and the time scale separation between the fast and slow variables. In particular, we are able to construct a scale adaptive scheme that requires minimal computational time for constructing a general parameterization scheme. 

At this regard, we have chosen to perform our analysis on the Lorenz '96 model \citep{Lorenz1996}. The Lorenz '96 model provides a conceptually meaningful yet extremely simplified representation of the atmosphere; there are two sets of variables, one describing the dynamics on large scale (so-called synoptic variables), and one characterizing the dynamics on small scale (so-called convective variables). The convective variables are divided in as many subgroups of equal size as the number of synoptic variables, each subgroup being coupled to a different synoptic scale variable. The system is then characterized by coupling within and across scales of motions. The Lorenz '96 model has quickly established itself as one of the reference models in nonlinear dynamics for testing e.g. data assimilation \citep{Trevisan2004,Trevisan2010}, schemes and properties of Lyapunov exponents and covariant Lyapunov vectors and is becoming increasingly popular also within the community of statistical mechanics \citep{Abramov2008,Hallerberg2010,Lucarini2011,Gallavotti2014}.

The evolution equations of the Lorenz '96 model can be written as:
\begin{equation} \label{eq:lor1before}
  \frac{dX_k}{dt}=X_{k-1}(X_{k+1}-X_{k-2})-X_k+F_1-\frac{hc}{b}\sum\limits_{j=1}^JY_{j,k},
\end{equation}
\begin{equation} \label{eq:lor2before}
  \frac{dY_{j,k}}{dt}=cbY_{j+1,k}(Y_{j-1,k}-Y_{j+2,k})-cY_{j,k}+\frac{hc}{b}X_k,
\end{equation}
with $k=1,...,K$; $j=1,...,J$. The boundary conditions are defined as 
\begin{equation}
\begin{split}
  X_{k-K}=X_{k+K}=X_k,\\
  Y_{j,k-K}=Y_{j,k+K}=Y_{j,k},\\
  Y_{j-J,k}=Y_{j,k-1},\\
  Y_{j+J,k}=Y_{j,k+1}.
\end{split}
\end{equation}
The latitudinal circle is divided into $K$ sectors, each one corresponding to one \textit{synoptic} slow $X$ variable. Each $X$ variable is coupled to $J$ \textit{convective} fast $Y$ variables. As discussed in detail later, the constant $c$ defines the time scale separation between the fast and slow variables (see also the general form of a multiscale system as given in Eqs \eqref{newperturbedX}-\eqref{newperturbedY}), while the amplitude of the fluctuations is determined by $b$, while $h$ controls the strength of the coupling.

In absence of forcing and dissipation, the sum of the squares of the variables (the \textit{energy} of the system) is conserved. For a detailed description of the statistical mechanical and conservation properties of the system (yet in a simplified version), the reader is encouraged to look into \cite{Lucarini2011,Blender2013,Gallavotti2014}.

We note that the coupling between the $X$ and the $Y$ terms has the simplified form discussed in the previous section (what we referred to as the independent coupling), and is linear. This simplifies the treatment below, which is nonetheless possible also for more complex forms of coupling.

The choice of the parameters defining the strength of the external forcing, the number of sectors and subsectors, the strength of the coupling, the relative amplitude of the fluctuations and the time scale separation between the two systems determines the properties of the dynamical system.
\newline The original parameters chosen by  Lorenz are $c=10.0$, $b=10.0$, $h=1.0$, $K=36$ and $J=10$ (providing therefore a total of $36$ $X$ variables and $360$ $Y$ variables). We remind that, following the original derivation of the model, $1$ unit of time is equivalent to $5$ days, while the usual integration time step is $0.005$, corresponding to $36$ minutes.

When one is well within the chaotic regime (e.g. $F_1$ is sufficiently large) and considers a sufficiently large number of sectors (and subsectors), it is reasonable to expect to be able to define intensive properties that are stable with respect to the specific choice of K and J, see discussion in \cite{Gallavotti2014} for a simpler version of the model.

\begin{figure}
  \includegraphics[width=\linewidth]{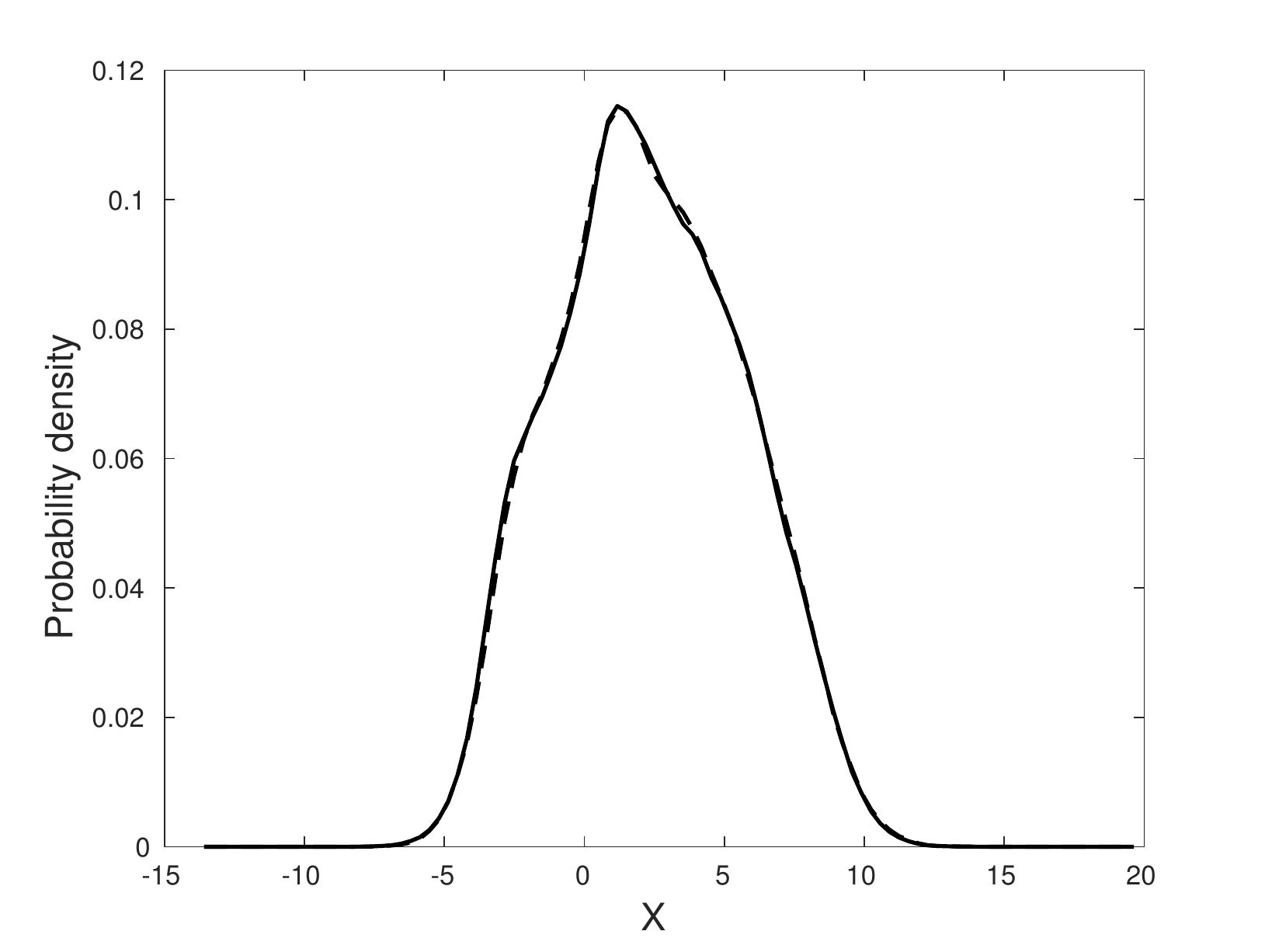}
  \caption{Probability density of the $X$ variable for the original (solid line) and the modified (dashed line) Lorenz 96 model. See text for details.}
  \label{fig:probdenscomp}
\end{figure}

\begin{figure}
  \includegraphics[width=\linewidth]{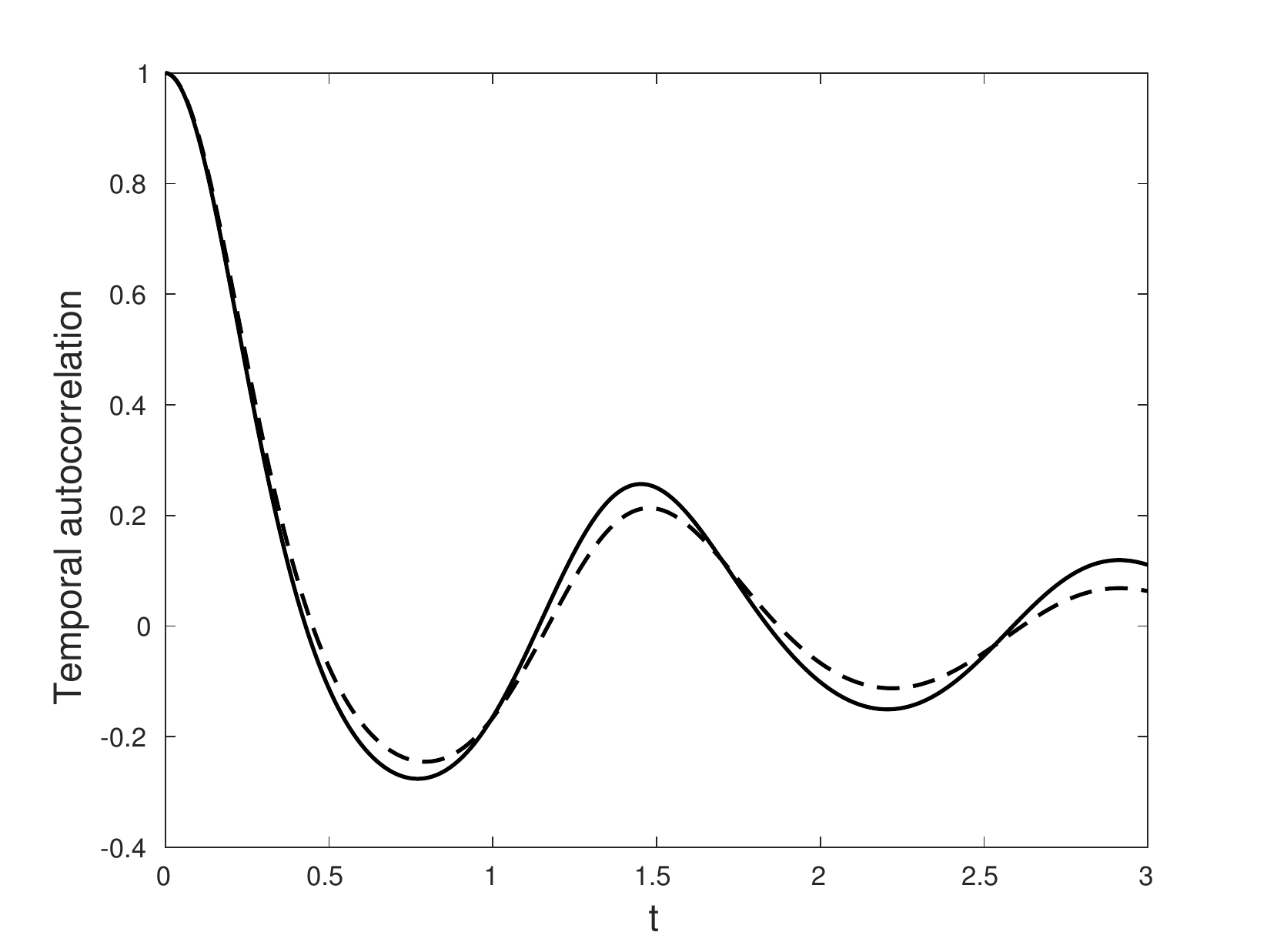}
  \caption{Time autocorrelation of the $X$ variable for the original (solid line) and the modified (dashed line) Lorenz 96 model. See text for details.}
  \label{fig:autocorrcomp}
\end{figure}

\begin{figure}
  \includegraphics[width=\linewidth]{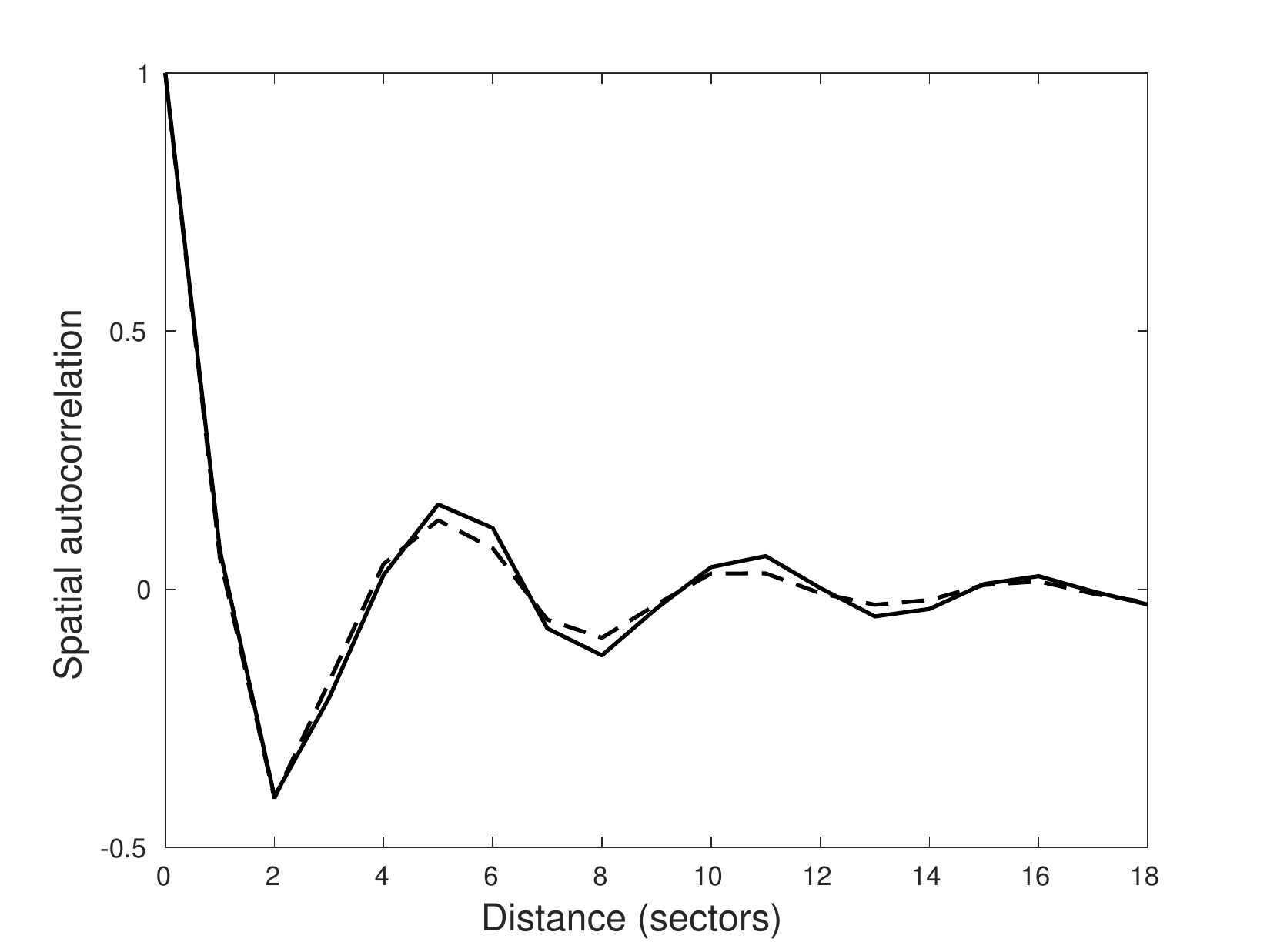}
  \caption{Spatial autocorrelation of the $X$ variable for the original (solid line) and the modified (dashed line) Lorenz 96 model. See text for details.}
  \label{fig:sautocorrcomp}
\end{figure}

We have implemented two modifications to the Lorenz '96 model:
\begin{itemize}
\item We have introduced a forcing term also in the equations describing the dynamics of the $Y$ variables (see Eq. \eqref{eq:lor2}), in order to represent the direct effect of forcings at small scales (mimicking, e.g., the impact of direct solar forcing on convective motions). This has the effect of making the fast variables an active component of the system: they  can also pump energy into the $X$ variables and are not exclusively dissipating energy coming from larger scales.
\item We have changed the boundary conditions on the $Y$ variables in such a way that the fast variables of different sectors do not interact with each other, in the spirit of having the fast variables representing sub-grid scale phenomena (see Eqs. \eqref{eq:newboundary}). Note that if $J\gg1$ and we are in a chaotic regime, it is reasonable that this change has negligible impact on the statistics of the system, as information does not propagate efficiently between convective variables belonging to neighbouring sectors. Additionally, the parameterization becomes easier to implement, because, following the basic idea behind super-parameterization, subgrid variables belonging to different $X$ sectors are independent and equivalent in the uncoupled case (see Eqs. \eqref{eq:unperturbedX}-\eqref{eq:unperturbedY}).
\end{itemize}
Therefore, the evolution equations \eqref{eq:lor1before}-\eqref{eq:lor2before} are modified as follows:
\begin{equation}  \label{eq:lor1}
  \frac{dX_k}{dt}=X_{k-1}(X_{k+1}-X_{k-2})-X_k+F_1-\frac{hc}{b}\sum\limits_{j=1}^JY_{j,k},
\end{equation}
\begin{equation}  \label{eq:lor2}
  \frac{dY_{j,k}}{dt}=cbY_{j+1,k}(Y_{j-1,k}-Y_{j+2,k})-cY_{j,k}+\frac{c}{b}F_2+\frac{hc}{b}X_k,
\end{equation}
with modified boundary conditions
\begin{equation}
\begin{split} \label{eq:newboundary}
  X_{k-K}=X_{k+K}=X_k,\\
  Y_{j-J,k}=Y_{j+J,k}=Y_{j,k}.
\end{split}
\end{equation}
The parameter $\epsilon$ in Eqs. \eqref{eq:perturbedX}-\eqref{eq:perturbedY} is $\frac{hc}{b}$ and the coupling terms are $\Psi_X=-\epsilon \sum\limits_{j=1}^JY_{j,k}$ and $\Psi_Y=\epsilon X_k$, $b$ defines the ratio between the typical size of the $X$ and $Y$ variables, while the parameter $\gamma$ controlling the scale separation is given by $c$. We choose $F_1=10.0$ and $F_2=6.0$, so that chaos is realized in the uncoupled version of the system (obtained from Eqs. \eqref{eq:lor1}-\eqref{eq:lor2} by setting $h=0$) for both the large and small scale variables of the system separately. {We choose for $h$, $b$, $c$, $K$, and $J$ the standard values mentioned above. We have verified that the change in the boundary conditions for the $Y$ variables has a negligible effect on the statistical properties of the $X$ variables: the pdf of each $X$ variable (Fig. \ref{fig:probdenscomp}), its time correlation (Fig. \ref{fig:autocorrcomp}), and the spatial correlation of the $X$ variables at zero time lag (Fig. \ref{fig:sautocorrcomp}) are virtually identical for the original and modified Lorenz '96 model.} The presence of chaos and of a corresponding nontrivial invariant measure for the $Y$ variables are necessary for being able to construct the W-L parameterization. In Appendix A we discuss how such a requirement can be relaxed through a suitable re-definition of the background around which the perturbative theory is applied.

We now show how to practically construct a scale-adaptive parameterization. This provides us with a great deal of flexibility and extremely parsimonious numerical costs. We show that the uncoupled evolution equation for the $Y$ variables (Eq. \eqref{eq:unperturbedY}) can be written in a universal form. In fact, it is  easy to check that, operating the substitutions
\begin{equation} \label{eq:rescc}
  \tau=ct
\end{equation}
and
\begin{equation} \label{eq:rescb}
  Z_{j,k}=bY_{j,k},
\end{equation}
we get {\color{black}for the uncoupled evolution equation for the rescaled Y variables:}
\begin{equation} \label{eq:rescaledY}
  \frac{dZ_{j,k}}{d\tau}=Z_{j+1,k}(Z_{j-1,k}-Z_{j+2,k})-Z_{j,k}+F_2 .
\end{equation}
Therefore, for all values of $h$, $b$, and $c$ we can construct the parameterizations just by resorting to the invariant measure of Eq. \eqref{eq:rescaledY} and adopting the suitable rescaling. Note that in the case of this specific system we are able to rescale also the size of the $Y$ variable and achieve a higher degree of flexibility than in the general case discussed above. This emphasizes the scale-adaptivity of the approach proposed here, and makes sure that only modest computation effort is needed to deal with the problem of parameterization.


Note that, compared to the general case of multiscale system discussed before, in this case we have the additional problem that changing the value of $c=\gamma$ leads also to an increase in the value of $\epsilon$, so that large values of $c$ might break the weak coupling hypothesis. The problem can be circumvented by increasing at the same time the value of $b$ or considering smaller values of $h$. 

\subsection{Constructing the Parameterization}
 
The first order term in the parameterization is recovered using ergodicity and averaging $D(X)$ in Eq. \eqref{eq:Dterm}. {\color{black}By symmetry, the coupling is the same for all the X variables:}
\begin{equation} \label{eq:m1}
  D_k(X_k)=D(X)=D=-\frac{1}{b} \lim_{T\to\infty}\frac{1}{T}\int_0^T \sum\limits_{j=1}^JZ_{j,k} (\tau)d\tau,
\end{equation}
where {\color{black}$k=1, \ldots, K$ and} the average is performed by integrating Eq. \eqref{eq:rescaledY}.  
\newline The value of this term is $\frac{-20.12}{b}$ for all $k'$s; choosing $b=10$ we get $D_k(X_k)=-2.012$. Therefore, the coupling between fast and slow scales leads on the average to a reduction in the effective forcing applied to the slow variables. In other terms, this indicates a net energy flux from slow to fast variables. {\color{black}Despite the simplicity of the model considered here and of the coupling between the X and Y variables, this corresponds to the effect of introducing eddy viscosity in more complex fluid dynamical models.}

\begin{figure}
  \centering
  \includegraphics[width=\linewidth]{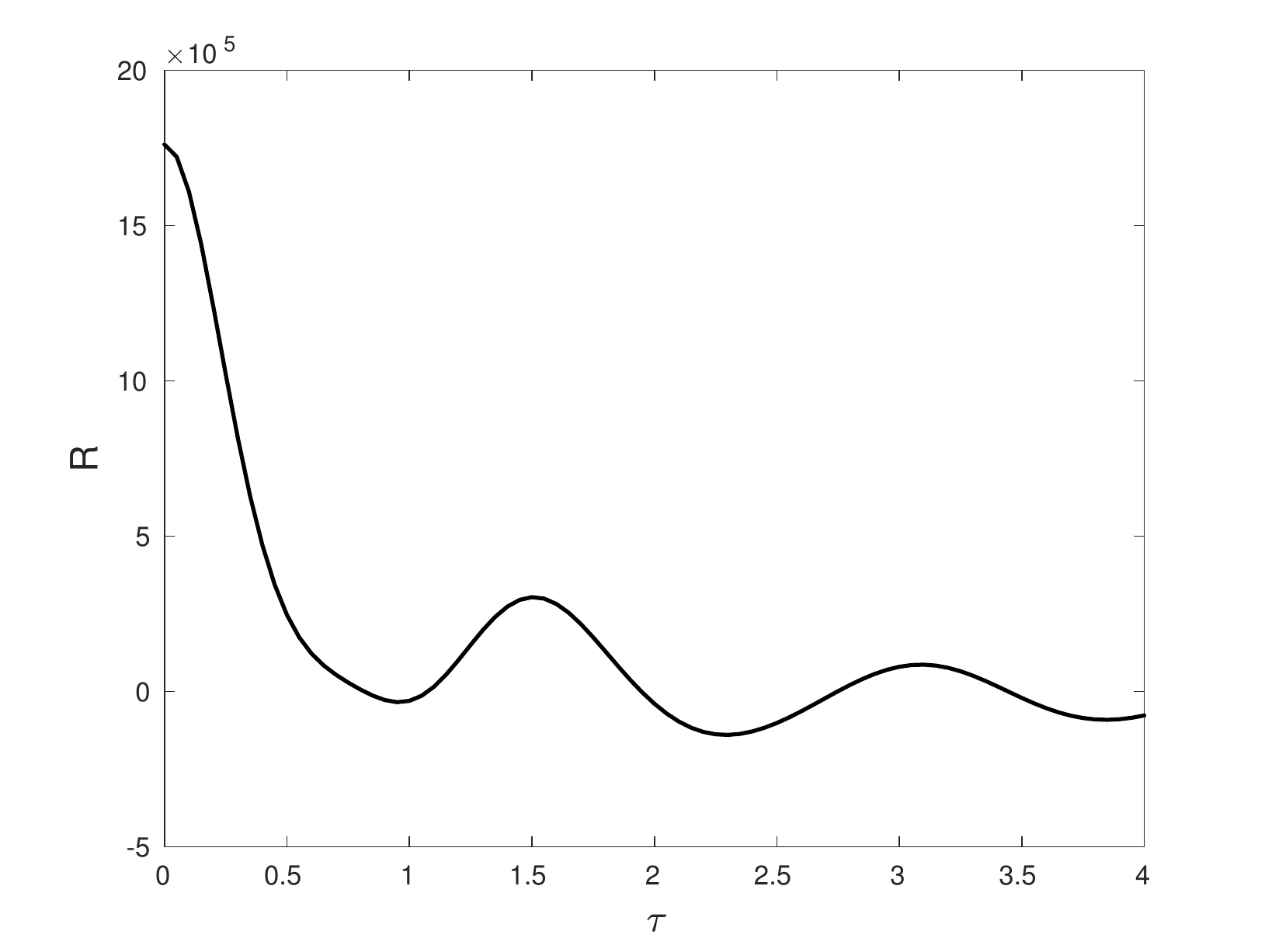}
  \caption{Time lagged autocovariance of the noise term $\sigma(t)$ with $b=10$ and $h=1$. See text for details.}
  \label{fig:m2ac}
\end{figure}

The $k^{th}$ component of the stochastic term $S\{X\}$ in Eq. \eqref{eq:m2} is constructed as an additive noise $\sigma(t)$ featuring the following lagged covariance:
\begin{equation} \label{eq:autocovariance}
  R_k(\tau)=R_k(ct)=\lim_{T\to\infty} \frac{1}{T} \int_0^T (-\sum\limits_{j=1}^J\frac{Z_{j,k}(\tau_1)}{b}-D)(-\sum\limits_{j=1}^J \frac{Z_{j,k}(\tau+\tau_1)}{b} -D) d\tau_1 ,
\end{equation}
where the evolution of the $Z$ variables is given by Eq. \eqref{eq:rescaledY} and the covariance is reported in Fig. \ref{fig:m2ac}.
We can construct surrogate time series of $\sigma$ to be used for the parameterized simulation either from properly resampling time series of the fluctuation term $-\sum\limits_{j=1}^J\frac{Z_{j,k}}{b}-D$ or by reproducing them using simple stochastic models like those belonging to the $AR(n)$ family. We follow the second route, taking advantage of the software package ARFIT \citep{Neumaier2001,Schneider2001}. {\color{black}Note that this term describes the backscatter of energy from the small towards the large scales.}
\newline Note that, as the argument of the function is $ct$, we have that in the limit of $c\to\infty$ the autocovariance tends to zero for all $t>0$, because the function $R$ tends to zero for large values of its argument, while one has for all values of $c$ that $R(0)$ is finite. As a result, one obtains as limit a white noise of vanishing amplitude {\color{black}for any fixed value of $\epsilon$.}   

We now wish to provide an explicit expression of the $k^{th}$ component of the non-markovian term $M\{X\}$ given in Eq. \eqref{eq:m3}. We express the memory kernel {\color{black}$h_k(\tau,\tilde{X_k})$} (where $\tau=ct$) as follows:
{\color{black}
\begin{equation} \label{eq:memorykernel}
  h_k(\tau_1,\tilde{X_k})= - \frac{1}{b} \tilde{X_k} H(\tau_1),
\end{equation} }
where
\begin{equation} \label{eq:kernelH}
  H(\tau_1)=lim_{\Omega\to\infty} \frac{1}{\Omega} \int_0^\Omega \sum\limits_{j=1}^J \frac{\partial}{\partial Z_{j,k}(\omega)} Z_{j,k}(\tau_1+\omega) d\omega .
\end{equation}

\begin{figure}
  \includegraphics[width=\linewidth]{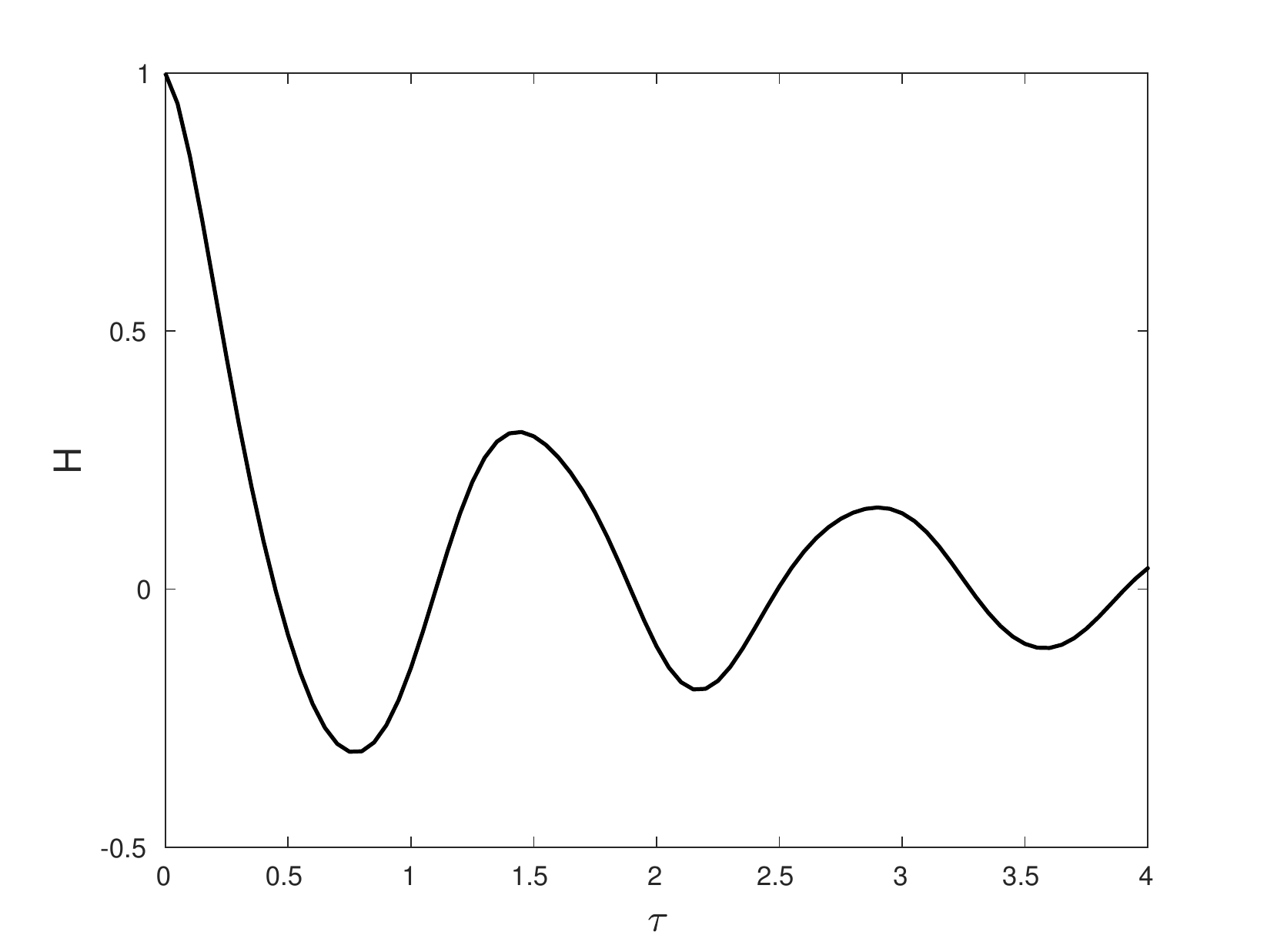}
  \caption{ Memory effects as measured by the factor $H$, see Eq. \eqref{eq:kernelH}, with $b=10$ and $h=1$. See text for details.}
  \label{fig:m3ws}
\end{figure}

In Fig. \ref{fig:m3ws} we plot the factor $H$ on the right hand side of Eq. \eqref{eq:memorykernel}, this clarifies that the kernel weighs less states of the $X$ variables with larger time separation, as expected. Increasing the value of $c$ leads to a compression of the time axis by a factor of $c$. Since {\color{black}$H(\tau)\to 0$} in the limit of $c\to\infty$, $h$ vanishes for all values of $t>0$. Hence, memory effects disappear in this limit.

We expect that, for a given value of $\epsilon$, the larger the value of $c$, the more dominant is the contribution to the parameterization coming from the deterministic first order term.

Note that \cite{Abramov2016} addresses the problem of parameterizing a modified version of the Lorenz'96 system similar to the one presented here by a modified version of the homogenization method. The derived parameterization is different from what obtained here as the homogenization method assumes infinite separation of scales between the fast and slow variables. Abramov obtains a stochastic contributions that is always white (yet its variance depends on the time scale separation), and an extra deterministic linear term that, from construction, might point at a surrogate way to implicitly deal with memory effects.

\section{Performance of the parameterization}

In this section a series of statistical tests are performed in order to check the skill of the W-L parameterization. In what follows, we refer to Eqs. \eqref{eq:lor1}-\eqref{eq:lor2} as the \textit{coupled model}. The \textit{uncoupled model} is, instead, given by the evolution equations of Eq. \eqref{eq:lor1} where the last term is excluded (or, equivalently, $h$ is set to 0). The model with \textit{first order parameterization} is obtained by inserting expression \eqref{eq:m1} in Eq. \eqref{eq:wl_par} and disregarding the other terms. The model with \textit{second order parameterization} is obtained by inserting in Eq. \eqref{eq:wl_par} both the first and second order terms. We test the skill of the parameterization in reproducing the statistical properties of the coupled model and compare it to the performance of the parameterization constructed according to the method proposed by \cite{Wilks2005}. We choose for this test the standard values of the parameters $c=10$, $b=10$, $h=1$; every other possible choice for these factors can be covered through a proper rescaling of the values for $D$, $S$ and $M$, as shown in sections 4.1 and 4.2.

Wilks proposed an empirical parameterization of the fast dynamical variables in the Lorenz '96 model. The idea is to fit the \textit{unresolved tendencies} of the $X$ variables (\textit{i.e.} the forcings terms written as a function of the $Y$ variables) using a polynomial regression in the form
\begin{equation} 
  g_U(X_k)=b_0+b_1 X_k+b_2 X_k^2+b_3 X_k^3+b_4 X_k^4+e_k,
\end{equation} 
where the $b$s are the regression coefficients, while $e_k$ is a stochastic function constructed according to the following AR(1) process:
\begin{equation} 
  e_k(t)=\phi e_k(t-\Delta)+\sigma_e(1-\phi^2)^{1/2} z_k(t),
\end{equation} 
written in term of the fitting parameters $\phi$ (lag-1 autocorrelation of $e_k$), $\sigma_e$ (standard deviation of the process $e_k$), where $z_k$ is a Gaussian uncorrelated process with zero mean and unitary variance. The parameterized system is then written as follows: 
\begin{equation} 
  \frac{dX_k}{dt}=X_{k-1}(X_{k+1}-X_{k-2})-X_k+F_1-g_U(X_k).
\end{equation} 

Note that in the case of Wilks's parameterization, all terms are markovian and there is no clear justification of why the stochastic residual is captured by an $AR(1)$ process, nor of why a 4th order polinomial is chosen. On the other side, the W-L parameterization provides a simple constant as deterministic term $D(X)$ (see Eq. \eqref{eq:m1}), which seems an oversimplification compared to the fourth order polynomial used by Wilks. This clarifies that the two approaches are rather different in nature. {\color{black}We remark that the framework for parameterizations for slow-fast system recently proposed by \cite{Wouters2017} and based on the Edgeworth expansion might provide a sound basis for justifying and possibly deriving explicitly closures structurally analogous to what proposed by Wilks for the Lorenz '96 system.}

We test the ability of the parameterizations in reproducing the probability density function of the variable $X_k$, the lagged temporal correlation $Corr(t)=\langle X_k X_k(t)\rangle$, and the spatial correlations at zero time lag $Sp(l)= \langle X_k X_{k+l}\rangle$.

\begin{figure}
  \centering
  \includegraphics[width=\linewidth]{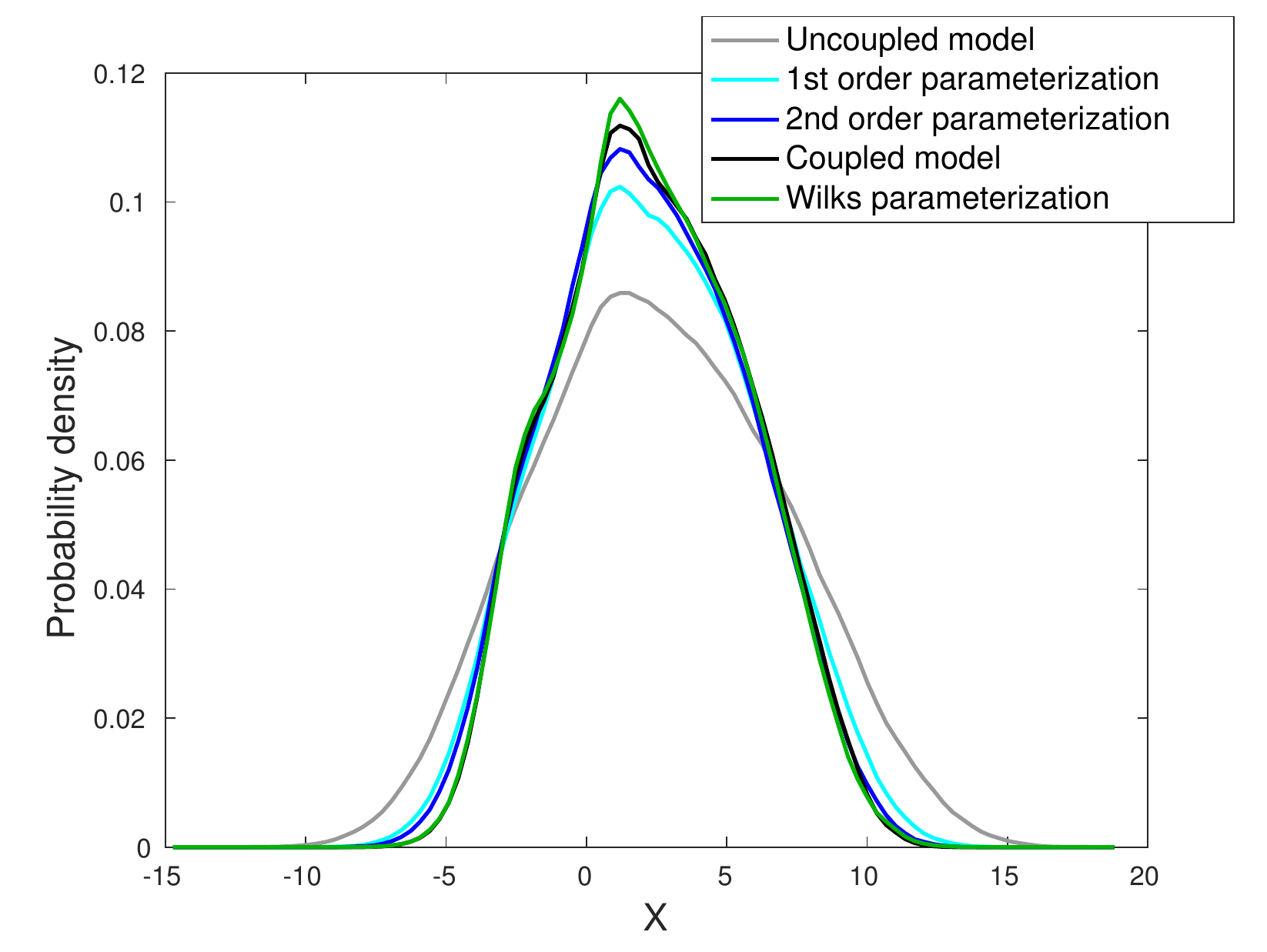}
  \caption{Probability density of the $X$ variable for the different models considered in the paper. See text for details.}
  \label{fig:prob_dens}
\end{figure}

Fig. \ref{fig:prob_dens} shows the probability density of the $X_k$ variables for all considered models. It is clear how the second order parameterization offers a better result with respect to the first order, which is in turn a clear improvement of the basic uncoupled system. Both Wilks's approach and the W-L parameterization provide rather good approximations of comparable quality for the distribution of the $X$ variable of the original system.

\begin{figure}
  \includegraphics[width=\linewidth]{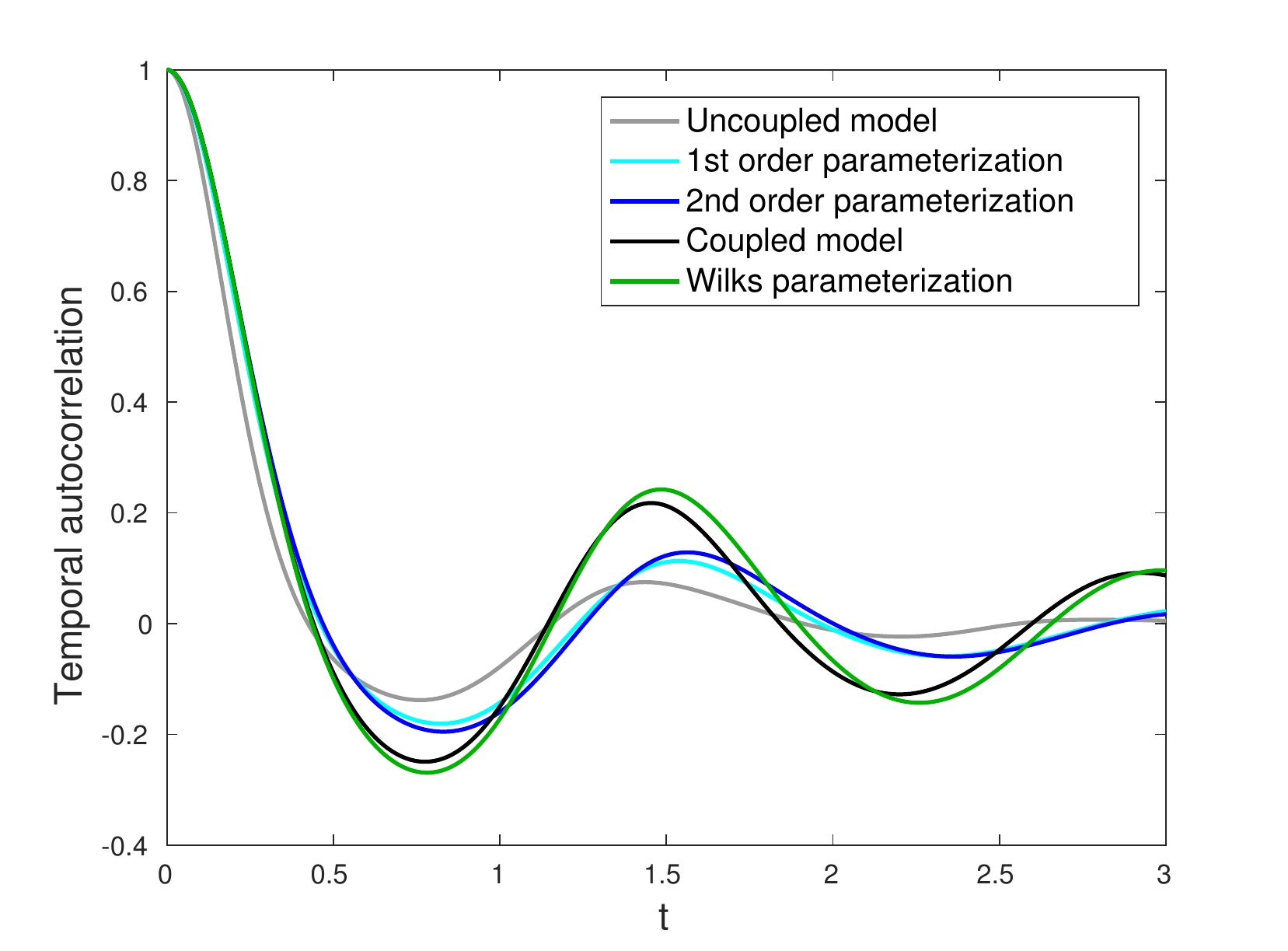}
  \caption{Temporal autocorrelation of the $X$ variable for the different models considered in the paper. See text for details.}
  \label{fig:temp_acorr}
\end{figure}

We then consider normalized second order properties for the $X$ variables. We first look at the lagged time autocorrelation (see Fig. \ref{fig:temp_acorr}). Higher order parameterizations lead to a better agreement with the coupled model, even if the improvement in the skill is most evident for small time lags. Nevertheless, the Wilks method provides very good results also for lags larger than 0.4 time units.

\begin{figure}
  \includegraphics[width=\linewidth]{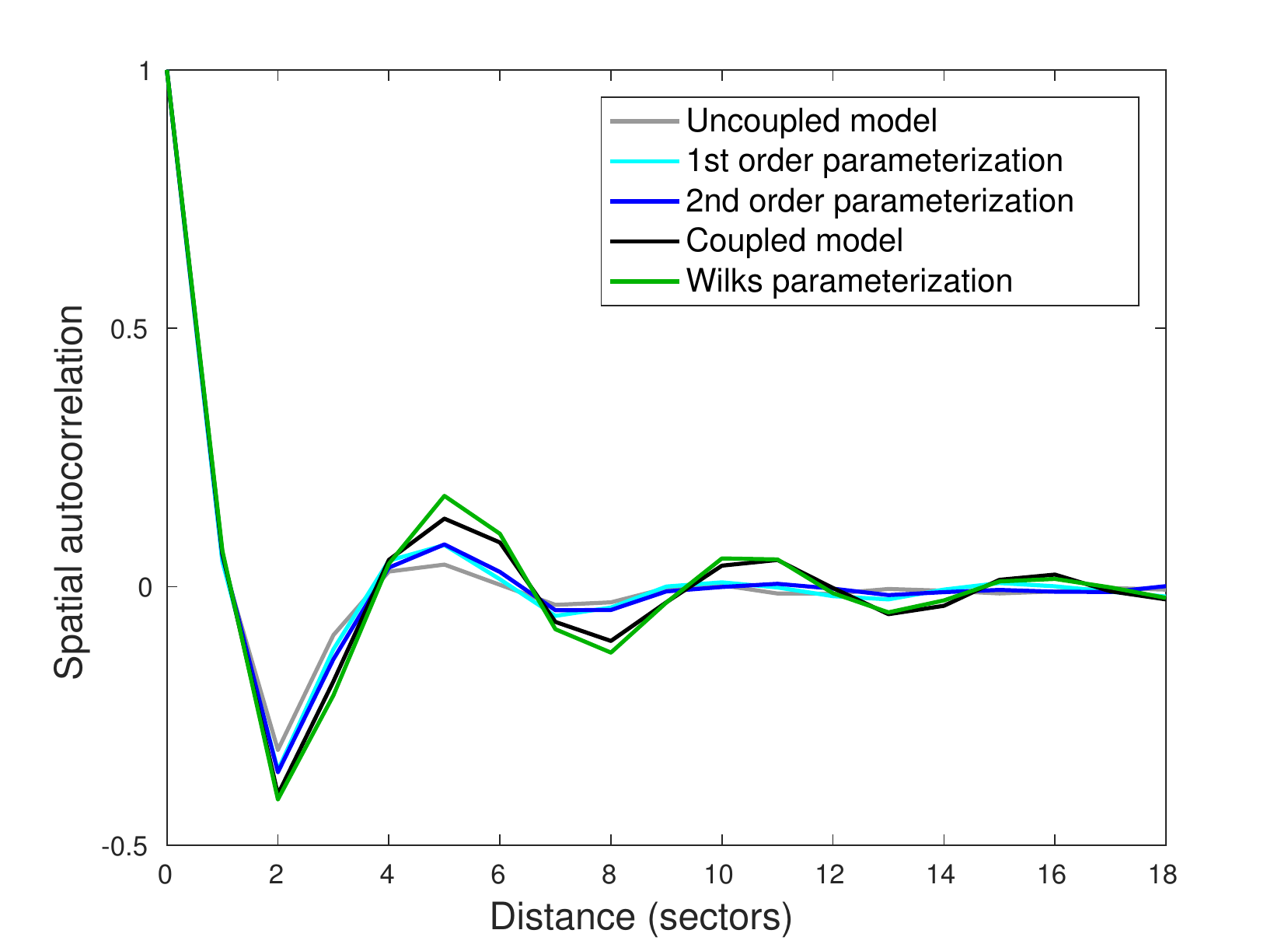}
  \caption{Spatial autocorrelation of the $X$ variable for the different models considered in the paper. See text for details.}
  \label{fig:sp_acorr}
\end{figure}

Fig. \ref{fig:sp_acorr} shows the performance of the parameterization in simulating the spatial correlation of the $X_k$ variable. We find that considering higher order approximations in the parameterizations we do not get a substantial improvement of the results, even if the first and second order parameterization lead to an improvement with respect to the uncoupled case. In this case Wilks's parameterization follows closely the full coupled model and overperforms the parameterizations constructed according to the method discussed here.

The analysis of the second and higher order moments is shown in the next section.

\subsection{Sensitivity to the Strength of the Coupling}

As our expansion is based on assuming the presence of weak coupling between the slow and the fast variables, it is crucial to test its performance as we vary the value of the coupling strength $h=\epsilon \frac{b}{c}$ (see Eqs. \ref{eq:lor1} and \ref{eq:lor2}) with respect to its standard value of 1. Note that we are treating the case where $b=c=10$ are held fixed, so that $h=\epsilon$ is changed in what follows. We look at the first moment and at the second, third and fourth central moments of the variable $X_k$. 

\begin{figure}
  \centering
  \includegraphics[width=\linewidth]{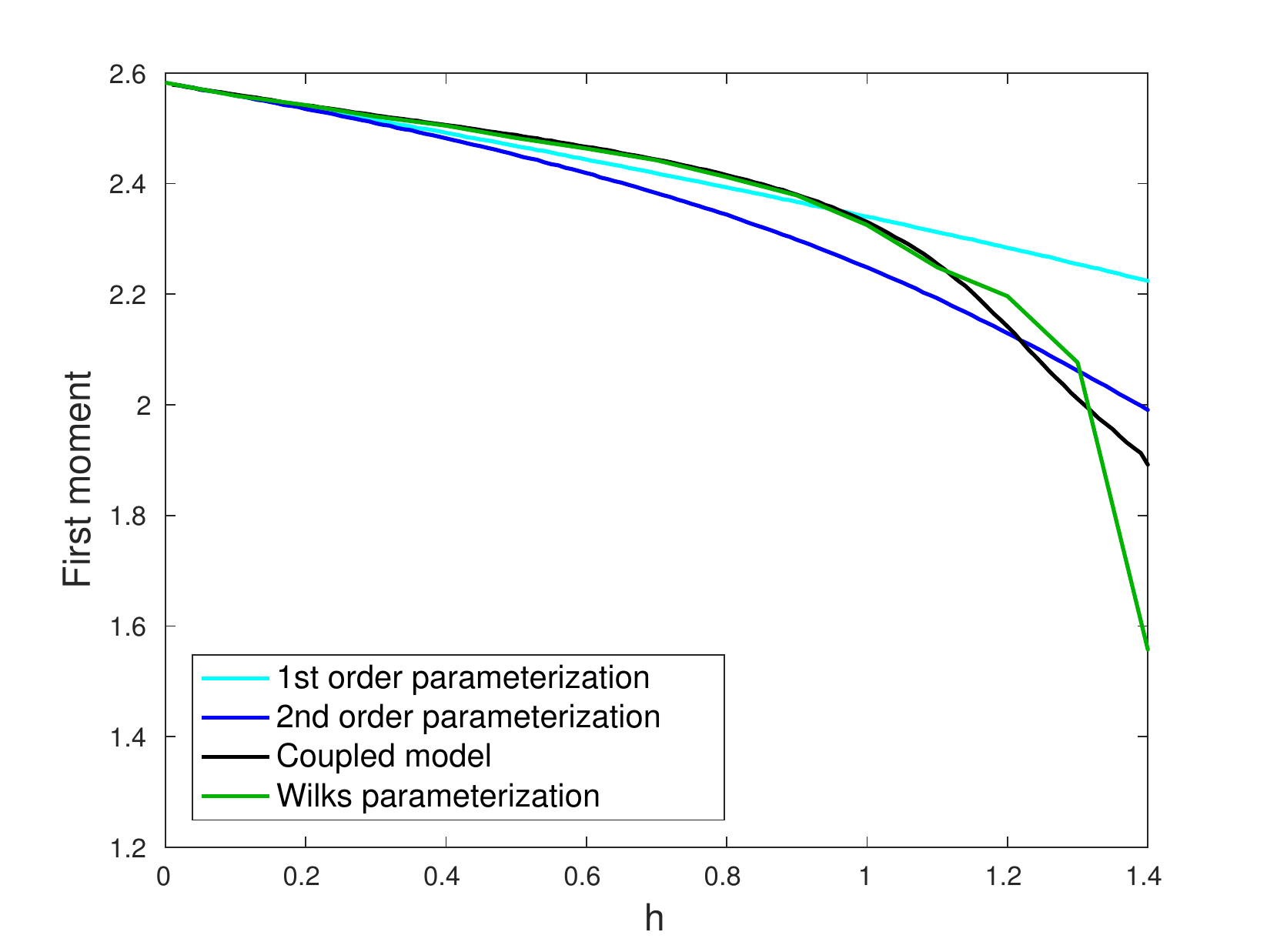}
  \caption{First moment as a function of the coupling strength for the different models considered in the paper. See text for details.}
  \label{fig:1stmom}
\end{figure}

Fig. \ref{fig:1stmom} shows that all parameterizations perform pretty well in terms of representing the first moment of $X_k$ for all considered values of $h<1.4$. Larger values of $h$ lead to a qualitative change in the properties of the system and fall outside the range of interest.

We note that, surprisingly, the first order parameterization constructed using the W-L method overperform the second order model for $h\lessapprox1$, which hints at the importance (at least in this case) of possibly developing a theory for the third order scheme, beyond the W-L parameterization.

\begin{figure}
  \includegraphics[width=\linewidth]{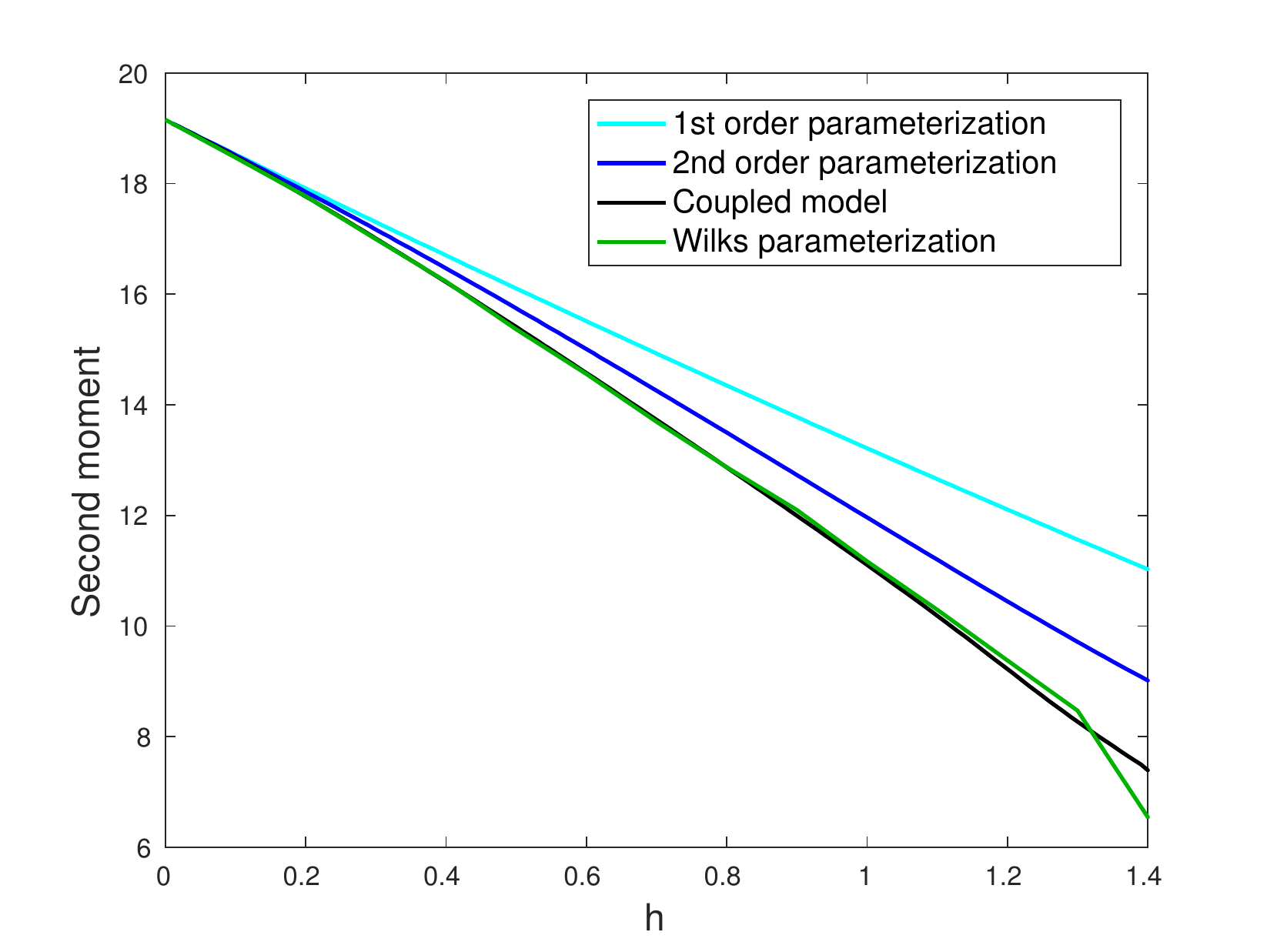}
  \caption{Second centered moment as a function of the coupling strength for the different models considered in the paper. See text for details.}
  \label{fig:2ndmom}
\end{figure}

\begin{figure}
  \includegraphics[width=\linewidth]{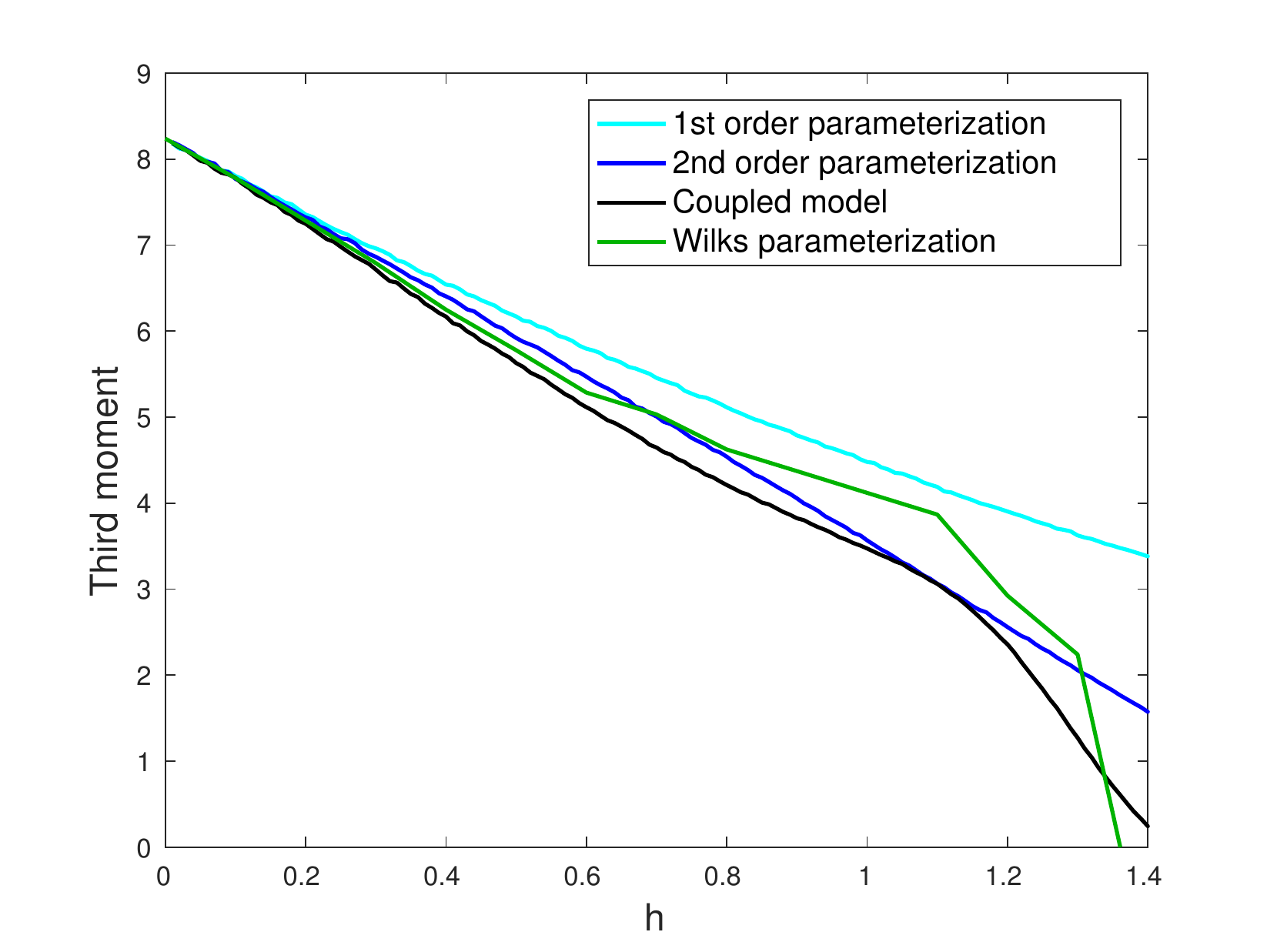}
  \caption{Third centered moment as a function of the coupling strength for the different models considered in the paper. See text for details.}
  \label{fig:3rdmom}
\end{figure}

\begin{figure}
  \includegraphics[width=\linewidth]{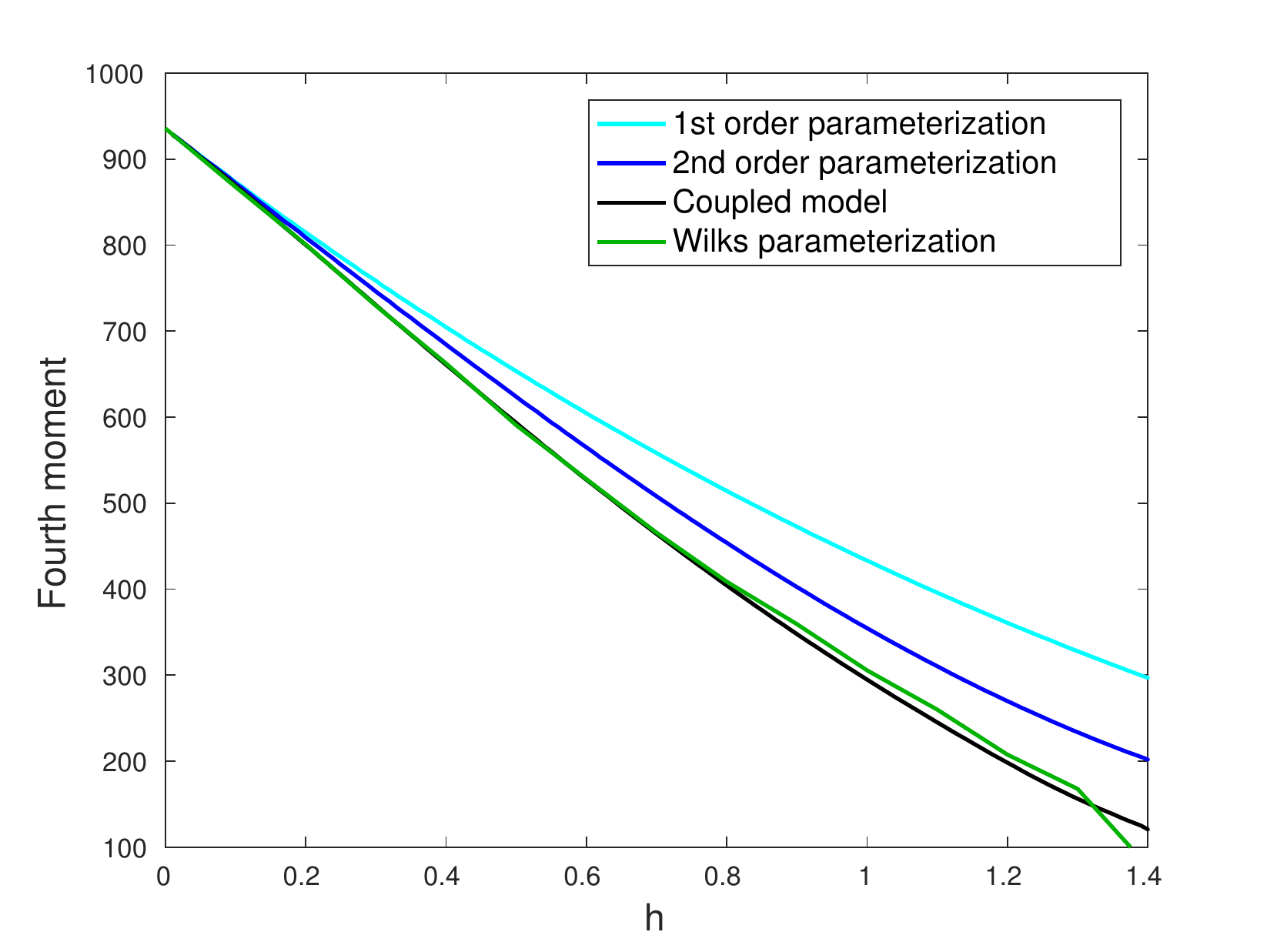}
  \caption{Fourth centered moment as a function of the coupling strength for the different models considered in the paper. See text for details.}
  \label{fig:4thmom}
\end{figure}

Figs. \ref{fig:2ndmom}, \ref{fig:3rdmom} and \ref{fig:4thmom} portray the performance of the parameterizations in reproducing the values of the second, third and fourth centered moments, respectively. We consistently find that, while all methods are quite successful, the Wilks parameterization provides the best results, with the second order model constructed with the W-L method coming close second.

We wish to underline that the Wilks parameterization needs to be constructed from scratch for each different value of $h$ (as well as of $b$ and $c$). This marks a fundamental difference with the parameterization tested in this study, where we need just to linearly rescale the first order term and quadratically rescale the two second order terms. Another problem shown by Wilks's method is the lack of stability in case of high values of $h$; as a matter of fact, in order to obtain results for $h>1.2$ we had to reduce drastically the time step in the numerical integration, thus having a much higher computational cost.


\subsection{Scale adaptivity}

The most relevant advantage of the W-L approach proposed here is {\color{black}that it allows one} to construct general parameterizations by suitably rescaling the three terms - deterministic, stochastic and non-markovian - after having estimated them through a single numerical simulation.
\newline The method proposed by Wilks is more precise for each given choice of the system's parameters but lacks such a flexibility, which might be of crucial relevance when trying to develop self-adaptive parameterizations. The coefficients appearing in the Wilks parameterizations (see Table \ref{tab:Wparameters}) cannot be readily predicted with suitable expressions.

\begin{table}
  \caption{Values of the constants determining the parameterization according to the Wilks method for various values of the model's parameter $c$.}
  \label{tab:Wparameters}
\noindent\makebox[\textwidth]{

  \begin{tabular}{l||c|c|c|c|c|c|r}
    $c$ & $b_0$ & $b_1$ & $b_2$ & $b_3$ & $b_4$ & $\sigma_e$ & $\phi$\\
    \hline
    $1$ & $1.694\times10^{-1}$ & $1.619\times10^{-2}$ & $7.507\times10^{-4}$ & $-7.868\times10^{-5}$ & $3.06\times10^{-7}$ & $1.04\times10^{-1}$ & $0.9995$\\
    $5$ & $9.122\times10^{-1}$ & $9.419\times10^{-2}$ & $-5.644\times10^{-3}$ & $4.025\times10^{-5}$ & $1.852\times10^{-5}$ & $4.659\times10^{-1}$ & $0.9996$\\
    $10$ & $1.81$ & $1.467\times10^{-1}$ & $-1.357\times10^{-3}$ & $1.446\times10^{-3}$ & $-1.313\times10^{-4}$ & $8.965\times10^{-1}$ & $0.9997$\\
    $20$ & $3.721$ & $4.861\times10^{-1}$ & $2.752\times10^{-2}$ & $-2.38\times10^{-2}$ & $2.427\times10^{-3}$ & $1.47$ & $0.9997$\\
    $100$ & $1.317\times10^{1}$ & $1.059$ & $2.969\times10^{-1}$ & $6.534\times10^{-2}$ & $3.421\times10^{-3}$ & $9.905\times10^{-2}$ & $0.9998$\\
  \end{tabular}} 
\end{table}

Using the general results for the first and second order terms of the W-L parameterization and adopting the suitable rescaling for the amplitude and the time axis discussed in the previous section, it is possible to explore an infinite range of scenarios for the values of $b$, $c$, and $h$. We present some examples below.

\begin{figure}
\begin{subfigure}{.5\textwidth}
  \centering
  \includegraphics[width=\linewidth]{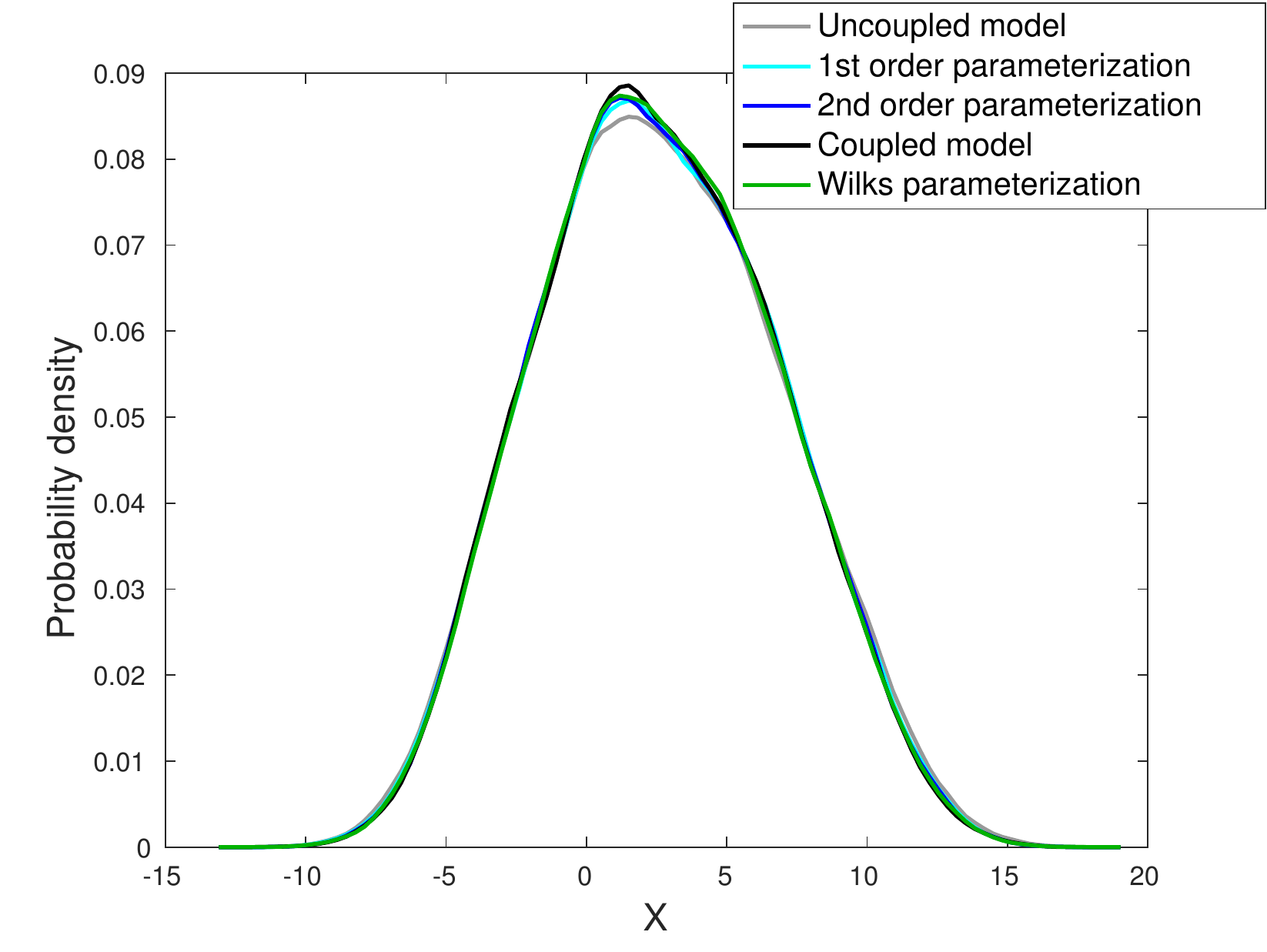}
  \caption{}
\end{subfigure}%
\begin{subfigure}{.5\textwidth}
  \includegraphics[width=\linewidth]{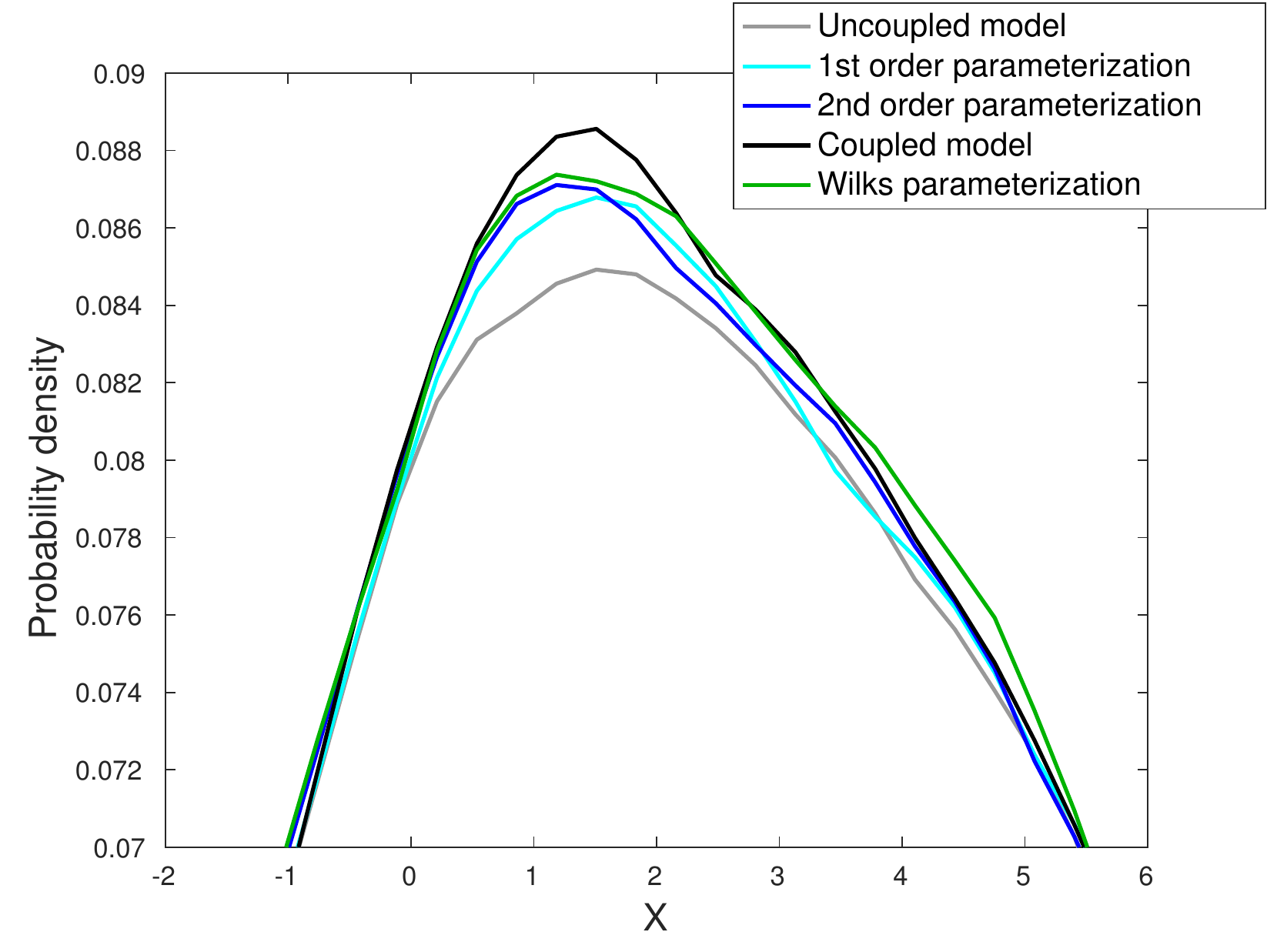}
  \caption{}
\end{subfigure}%
  \caption{a) Probability density of the $X$ variable in the case of $c=1$, $b=10$ and $h=1$. See text for details. b) Zoom on the peak of the distribution.}
  \label{fig:c1pd}
\end{figure}

In Fig. \ref{fig:c1pd} we show that the probability densities of $X_k$ obtained through the different parameterizations are in good agreement with what shown by the coupled model. 
Note that choosing $c=1$ implies also assuming that there is \textit{no} scale separation between the $X$ and the $Y$ variables. In fact, as discussed before, the W-L method can be used also in this case. 

\begin{figure}
\begin{subfigure}{.5\textwidth}
  \centering
  \includegraphics[width=\linewidth]{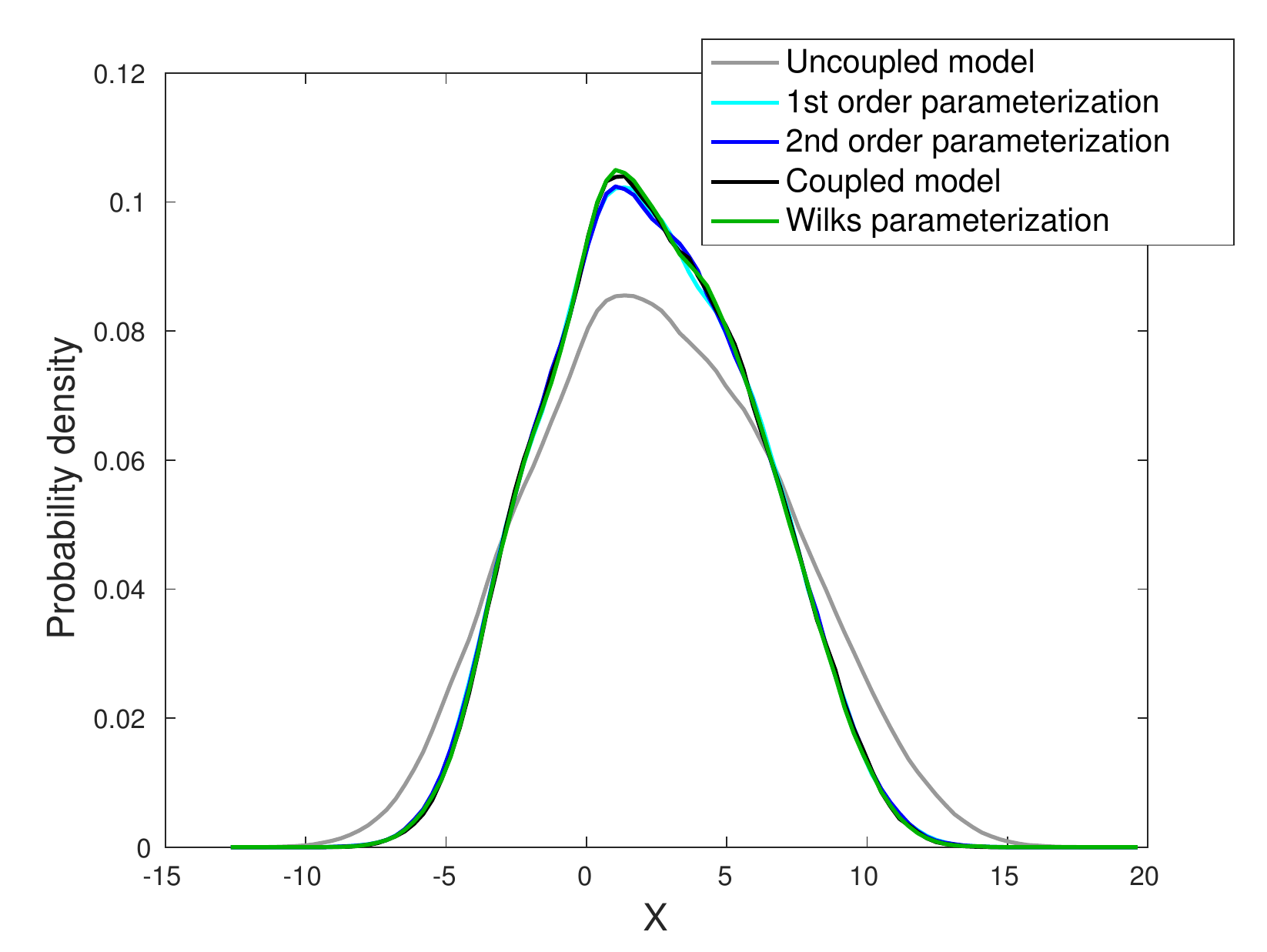}
  \caption{}
\end{subfigure}%
\begin{subfigure}{.5\textwidth}
  \includegraphics[width=\linewidth]{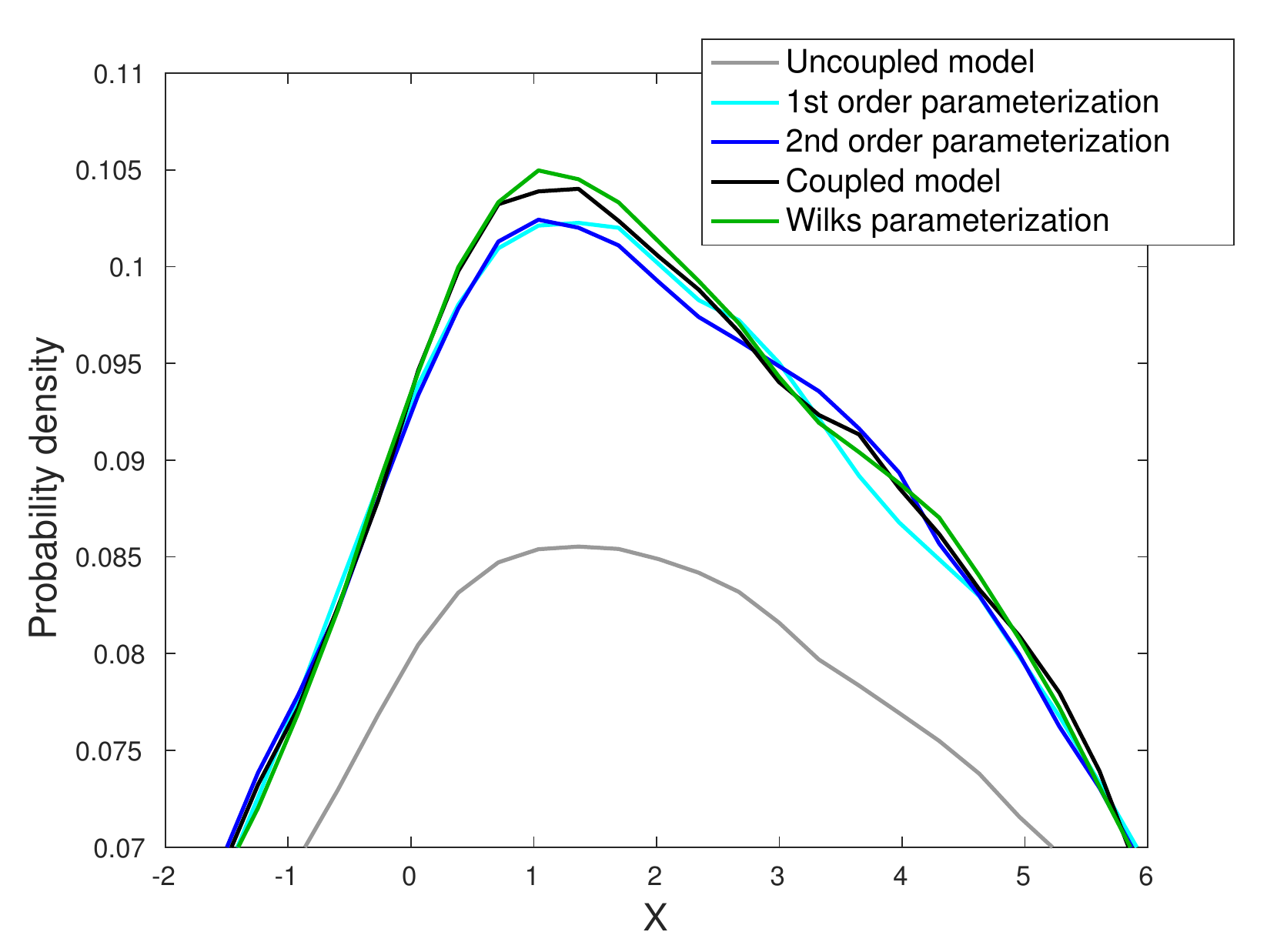}
  \caption{}
\end{subfigure}%
  \caption{a) Probability density of the $X$ variable in the case of $c=100$, $b=10$ and $h=0.1$. See text for details. b) Zoom on the peak of the distribution.}
  \label{fig:c100h1pd}
\end{figure}

Since we are treating the case where the coupling should not be too strong compared to the unperturbed vector flow (this is the condition under which we can use the W-L method), as said before, increasing the value of $c$ can be problematic unless we reduce accordingly the value of $h$ (or increase the value of $b$). We then show in Fig. \ref{fig:c100h1pd} the probability density function of the $X$ variable in the case $c=100$, $b=10$, $h=0.1$, with a much stronger coupling than the previous case of Fig. \ref{fig:c1pd}.
\newline In this case, it is clear that considering a parameterization is crucial for reproducing satisfactorily the statistics of the $X$ variable, and we observe that the first order parameterization is already rather successful. Note that as $c$ becomes larger, the memory term has a less and less relevant role and the stochastic contributions is rather similar to a white noise forcing.

\begin{figure}
  \includegraphics[width=\linewidth]{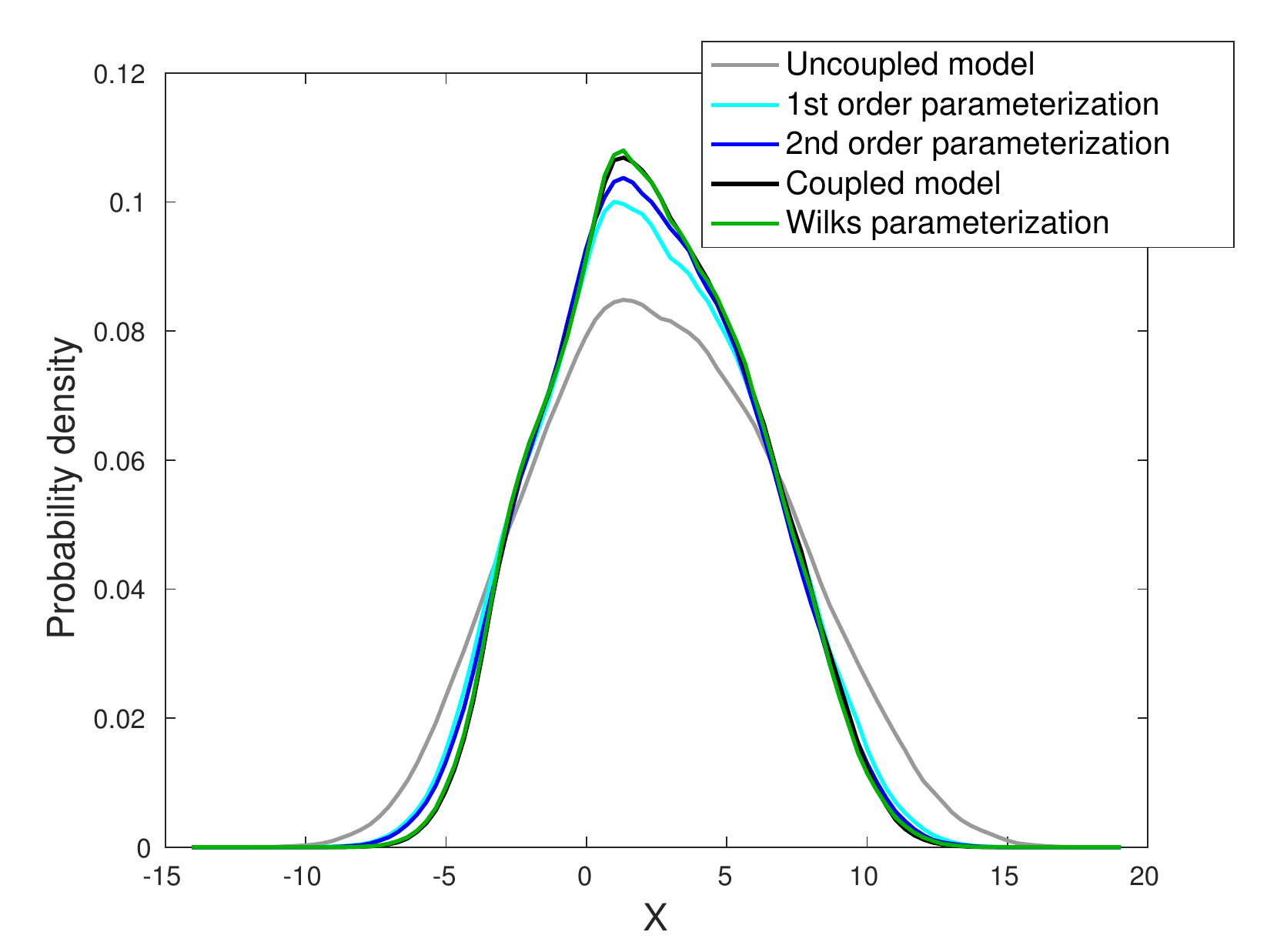}
  \caption{Probability density of the $X$ variable in the case of $c=5$, $b=8$ and $h=1.1$. See text for details.}
  \label{fig:stresspd}
\end{figure}

As last test (Fig. \ref{fig:stresspd}) we stress the rescaling of the model applying the transformation to all the parameters at the same time, shifting from the $c=10$, $b=10$, $h=1$ to the $c=5$, $b=8$, $h=1.1$ scenario.

\section{Conclusions}
Constructing accurate and efficient parameterizations is a central task in the numerical modelling of geophysical fluids, because these systems are characterized by variability on a large range of spatial and temporal scales. Parameterizations are expected to be able to improve our ability to predict and represent the statistical properties of the slow, large scales variables of interest bypassing the need for representing explicitly the dynamics of fast, small scale variables. Modern climate and weather prediction systems devote relevant resources to improving the parameterizations of subgrid scale processes. Unfortunately, parameterizations are usually constructed ad-hoc, targeting the optimisations of one of few specific properties of interest, and must usually be re-tuned each time the resolution of the model is changed or new components are added. This creates intrinsic uncertainties in the performance of the model and reduces its overall robustness. 

General mathematical findings suggest that parameterizations should include deterministic, stochastic, and non-markovian contributions \citep{Chekroun2015a,Chekroun2015b,Wouters2012,Wouters2013,Wouters2016}. In particular, non-white noise and non-markovian terms result from the finiteness of the scale separation between resolved and unresolved scales. Indeed, current developments in meteorology and climate science are strongly proposing supplementing the common deterministic parameterizations with stochastic components and satisfactory improvements in the skill are observed \citep{Palmer2008,Franzke2015,Berner2016}.

In this paper we implement the parameterization scheme developed for general systems by Wouters and Lucarini through a re-elaboration of the Ruelle response theory \citep{Ruelle1998,Ruelle2009} and, independently, through an expansion of the Mori-Zwanzig projection operator \citep{Zwanzig1960,Zwanzig1961,Mori1974}, where the coupling between the variables of interest and the variables we want to parameterize is seen as small perturbation to the uncoupled dynamics of the former ones, thus taking a weak coupling hypothesis. This parameterization describing  the dynamical impact of the neglected variables can be written as the sum of as a deterministic term (mean field effect) stochastic term (impacts of fluctuations), and non-markovian term (role of memory). We underline that following this point of view, instead, no hypotheses are taken on the scale separation between the systems, as opposed to well-exploited approaches as the homogenization method \citep{Pavliotis2008}. We also show that the W-L method can be used in general for constructing scale-adaptive parameterizations when multi-scale systems are considered.

We test this parameterization scheme on a mildly modified version of the Lorenz '96 two-level model \citep{Lorenz1996}, which is a prototypical multi-scale system of great interest for nonlinear science in general.  

We construct a scale adaptive parameterization able to describe accurately the coupling between slow and fast scales, to describe conditions of finite scale separation and to reach the infinite time separation limit.

In particular, we are able to construct explicitly the properties of the noise term responsible for the stochastic component of the parameterization and the memory kernel responsible for the non-markovian term.

The parameterization does a very good job in surrogating the effects of the fast variables, as tested by evaluating the expectation value and the correlation properties of the slow variables, and shows a great deal of flexibility when different intensity for the coupling strength is varied. 

We have also tested the parameterization discussed here against the heuristic approach previously proposed by \cite{Wilks2005}. The Wilks method allows for constructing detailed parameterizations for each choice of the systems parameters, and outperforms the parameterizations constructed following Wouters and Lucarini. Nonetheless, the Wilks parameterization is not scale adaptive and needs to be retuned each time we change one or more parameters of the system, whereas the W-L parameterization is universal {\color{black}within the approximation defined by Eq.\eqref{eq:rhoOthird}}, except for the application of an algebraic rescaling, as proved by our last tests. We argue that, depending on the specific problem one needs to address, an accurate ad-hoc method or the flexible but less precise method proposed here might prove more advantageous.

The flexibility of this approach has been demonstrated by changing by two orders of magnitude the time scale separation and also in the most general case when all the parameters $c$, $b$ and $h$ are changed with respect to the original values. It would be interesting to sistematically compare the parameterization described here with what recently proposed by {\color{black}\cite{Abramov2016} and \cite{Wouters2017} through modifications} of the homogenization method, in order to assess benefits and pitfalls of each approach. {\color{black}It would also be extremely interesting to test, for a given system of interest, the correspondences and differences between the point of view proposed here and the bottom-up, data-based, complementary approach proposed by \cite{Kondrashov2017}, which also allows for dealing with non-markovian effects. Especially promising seems the possibility of testing and comparing the scale-adaptivity of the two approaches.}

We wish to test next the relative performance of the parameterizations described here in terms of prediction of the state of the system described by the $X$ variables. Additionally, we plan to test sistematically what is presented in Appendix A, i.e. the flexibility we have in the theory used here of selecting different background states for constructing the parameterization.

What we have shown in this paper is, evidently, mostly a proof of concept aimed at essaying the potential (and the pitfalls) of the W-L approach {\color{black}and showing its scale-adaptivity, which had not been thoroughly studied before.} This is, {\color{black}together with the recent contributions by \cite{Wouters2016} and \cite{Demaeyer2015},} just the first step in the direction of understanding its applicability in GFD systems of practical interest, where large datasets need to be processed and the computation of the memory term seems at first sight problematic. In particular, we will aim at constructing filters for large eddy simulations \citep{Pope2004,Arakawa2004}. This clearly seems to be an ambitions task and further investigations are needed in this direction. {\color{black}In fact, the potential of the WL parameterization might be higher than what shown until now. In a recent publication, \cite{Lucarini2017} showed that it is possible to derive explicit formulas allowing for projecting imposed changes in the dynamics of the full system due to perturbations onto the reduced, parameterized dynamics. This paves potentially the way for constructing extremely flexible parameterizations. This is another direction of work worth investigating.}

\section*{Acknowledgement}
The authors wish to thank Sebastian Schubert for various fruitful discussions. GV was supported by the Hans Ertel Center for Weather Research (HErZ), a collaborative project involving universities across Germany, the Deutscher Wetterdienst and funded by the BMVI (Federal Ministry of Transport and Digital Infrastructure, Germany). VL acknowledges the financial support provided by the DFG cluster of excellence CliSAP and by the SFB/Transregio Project TRR181.

\appendix

\section{Forcing in the fast dynamics}

As discussed in \cite{Wouters2012,Wouters2013,Wouters2016}, a basic requirement for the proposed approach to allow for the construction of a parameterization for the Y variables is to have that the uncoupled dynamics of the $Y$ variables given in Eq. \eqref{eq:unperturbedY} features a non-trivial invariant measure and fast decay of correlations due to the presence of chaos. Physically, this requires presence of an external forcing  leading to the injection of energy for the $Y$ variables; this is achieved in the system studied here by choosing a sufficiently large value for the constant $F_2$. Another way to address such a  problem is shown in \cite{Wouters2016a}, where a stochastic forcing, corresponding to the presence of energy injection coming from even smaller, unresolved scale, is considered. 

In order to extent the method to physical situations where energy is injected only in the $X$ variables, we need to resort to a simple mathematical trick that amounts to changing the background state around which the perturbation induced by the presence of coupling is considered.

The idea is to rewrite Eq. \eqref{eq:perturbedY} {\color{black}(we refer here for simplicity to the case with $\Psi_X(X,Y)=\Psi_X(Y)$ and $\Psi_Y(X,Y)=\Psi_Y(X)$)} as follows:
{\color{black}
\begin{equation}
    \frac{dY}{dt}=F_Y(Y)+\epsilon G+\epsilon\Psi_Y(X)-\epsilon G ,
\end{equation} }

such that the vector flows defining the uncoupled dynamics and the coupling are defined as follows:
{\color{black}
\begin{align}
  & \widetilde F_Y(Y) = F_Y(Y)+\epsilon G ,\label{plusG}\\
  & \widetilde {\epsilon\Psi_Y}(X) = \epsilon\Psi_Y(X)-\epsilon G. \label{minusG}
\end{align} }

The choice of the \textit{artificial} forcing $G$ gives us a degree of flexibility and must obey only the requirement that $\dot{Y}=\tilde{F}_Y(Y)$ is chaotic. Note that, within the radius of expansion ensuring the validity of the {\color{black}perturbative} approach the specific choice of $G$ affects only weakly our final result. 

An obvious choice is to choose $G=\rho_{0,X}(\Psi_Y(X))$, which makes sure that, at zero order, the uncoupled system has a nontrivial dynamics, because we have chosen a background state where the Y variables receive from the $X$ variables approximately as much energy as in the fully coupled case. 

\begin{figure}
  \includegraphics[width=\linewidth]{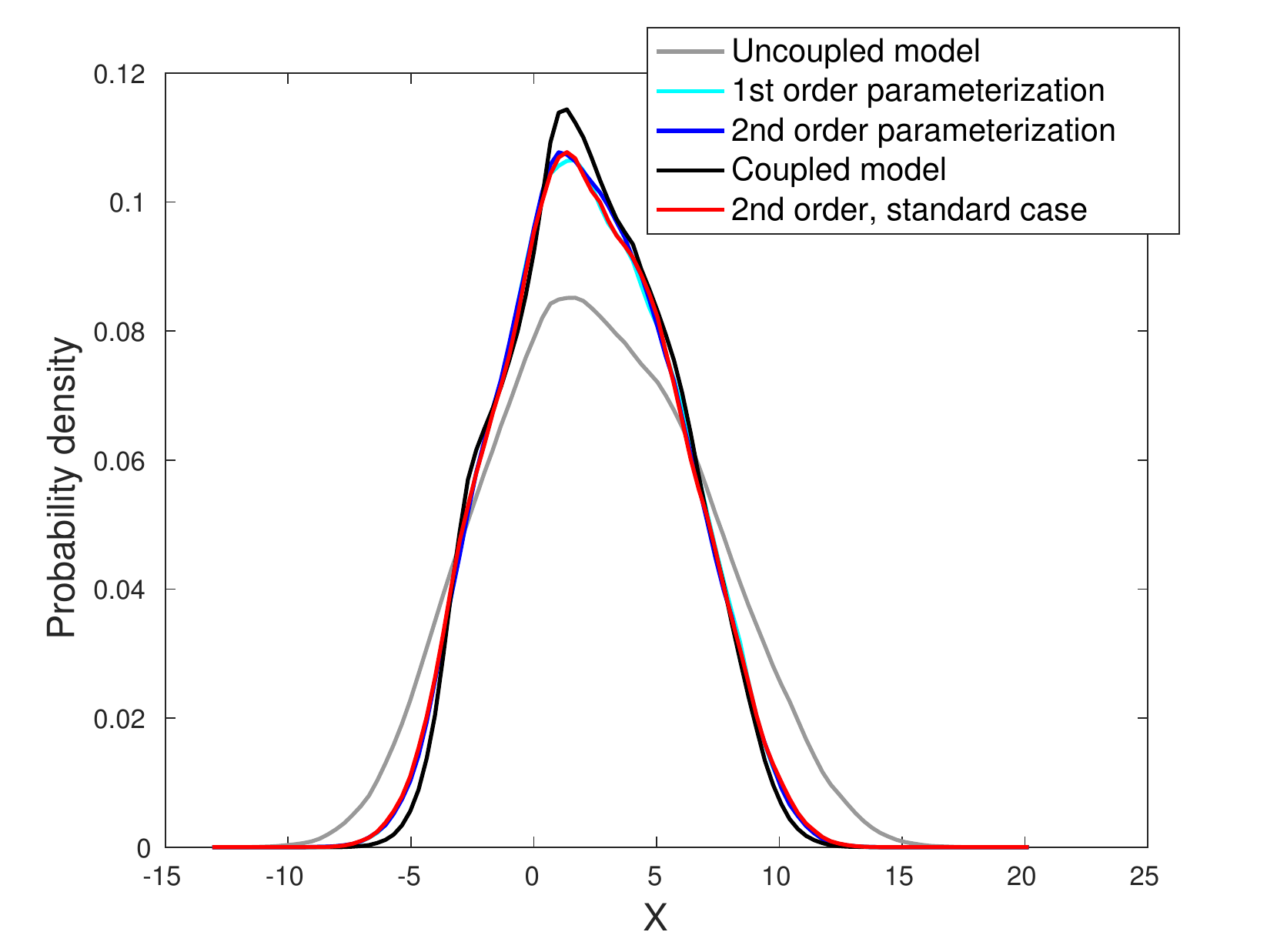}
  \caption{Probability density of the $X$ variable calculated adding $G$ to uncoupled $Y$ equation. The standard case is the one shown in section 4. See text for details.}
  \label{fig:probdensg}
\end{figure}

\begin{figure}
  \includegraphics[width=\linewidth]{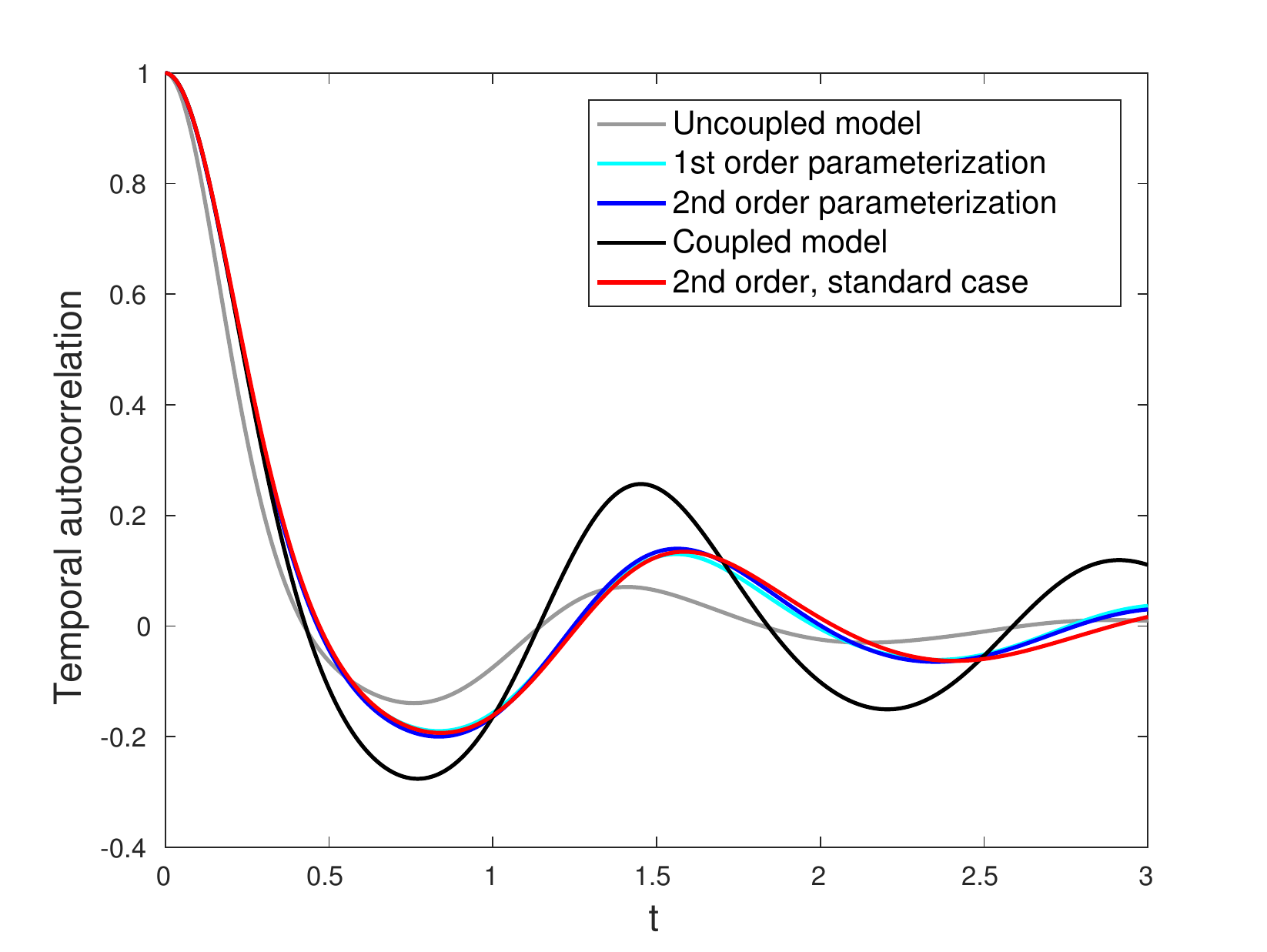}
  \caption{Time autocorrelation of the $X$ variable calculated adding $G$ to uncoupled $Y$ equation. The standard case is the one shown in section 4. See text for details.}
  \label{fig:autocorrg}
\end{figure}

The procedure can be repeated also in the case, like the one analyzed here, where we do not have the requirement of shifting the background state, and the natural definition of the uncoupled dynamics of the $Y$ variables given in Eq. \eqref{eq:unperturbedY} can be used. We have tested here this hypothesis by using the framework given in Eqs. \eqref{plusG}-\eqref{minusG} and choosing the standard values for the system's parameters and $G=\rho_{0,X}(\Psi_Y(X))=2.57$. In Figs. \ref{fig:probdensg} and \ref{fig:autocorrg} we show that, at the second order, we obtain almost undistinguishable results with respect to what shown in Fig. \ref{fig:prob_dens} for probability density and Fig. \ref{fig:temp_acorr} for the time autocorrelation using $F_2=6$ and $\frac{c}{b}F_2+G=8.57$ as forcing of the uncoupled $Y$ equation. 

\begin{thebibliography}{}

\bibitem[Abramov, 2016]{Abramov2016}
Abramov, R.~V. (2016).
\newblock {A Simple Stochastic Parameterization for Reduced Models of
  Multiscale Dynamics}.
\newblock {\em Fluids}, 1, 2(October 2015):1--18.

\bibitem[Abramov and Majda, 2008]{Abramov2008}
Abramov, R.~V. and Majda, A.~J. (2008).
\newblock {New approximations and tests of linear fluctuation-response for
  chaotic nonlinear forced-dissipative dynamical systems}.
\newblock {\em Journal of Nonlinear Science}, 18(3):303--341.

\bibitem[Arakawa, 2004]{Arakawa2004}
Arakawa, A. (2004).
\newblock {The cumulus parameterization problem: Past, present, and future}.
\newblock {\em Journal of Climate}, 17(13):2493--2525.

\bibitem[Arakawa et~al., 2011]{Arakawa2011}
Arakawa, A., Jung, J., and Wu, C. (2011).
\newblock {and Physics Toward unification of the multiscale modeling of the
  atmosphere}.
\newblock {\em Atmos. Chem. Phys.}, 11:3731--3742.

\bibitem[Berner et~al., 2016]{Berner2016}
Berner, J., Achatz, U., Batt{\'e}, L., Bengtsson, L., C{\'a}mara, A. D.~L.,
  Christensen, H.~M., Colangeli, M., Coleman, D. R.~B., Crommelin, D.,
  Dolaptchiev, S.~I., Franzke, C.~L., Friederichs, P., Imkeller, P.,
  J{\"a}rvinen, H., Juricke, S., Kitsios, V., Lott, F., Lucarini, V., Mahajan,
  S., Palmer, T.~N., Penland, C., Sakradzija, M., Storch, J.-S.~V., Weisheimer,
  A., Weniger, M., Williams, P.~D., and Yano, J.-I. (2016).
\newblock {Stochastic Parameterization: Towards a new view of Weather and
  Climate Models}.
\newblock {\em Bulletin of the American Meteorological Society}.

\bibitem[Blender and Lucarini, 2013]{Blender2013}
Blender, R. and Lucarini, V. (2013).
\newblock {Nambu representation of an extended Lorenz model with viscous
  heating}.
\newblock {\em Physica D: Nonlinear Phenomena}, 243(1):86--91.

\bibitem[Chekroun et~al., 2015a]{Chekroun2015a}
Chekroun, M.~D., Liu, H., and Wang, S. (2015a).
\newblock {\em {Approximation of Stochastic Invariant Manifolds}}.
\newblock {SpringerBriefs in Mathematics}. Springer International Publishing,
  Cham.

\bibitem[Chekroun et~al., 2015b]{Chekroun2015b}
Chekroun, M.~D., Liu, H., and Wang, S. (2015b).
\newblock {\em {Stochastic Parameterizing Manifolds and Non-Markovian Reduced
  Equations}}.
\newblock {SpringerBriefs in Mathematics}. Springer International Publishing,
  Cham.

\bibitem[Demaeyer and Vannitsem, 2017]{Demaeyer2015}
Demaeyer, J. and Vannitsem, S. (2017).
\newblock {Stochastic parameterization of subgrid-scale processes in coupled
  ocean-atmosphere systems: Benefits and limitations of response theory}.
\newblock {\em Quarterly Journal of the Royal Meteorological Society},
  143(703):881--896.

\bibitem[Eckmann and Ruelle, 1985]{Eckmann1985}
Eckmann, J. and Ruelle, D. (1985).
\newblock {Ergodic theory of chaos and strange attractors}.
\newblock {\em Reviews of Modern Physics}, 57(4):617--656.

\bibitem[Franzke et~al., 2005]{Franzke2005}
Franzke, C., Majda, A.~J., and Vanden-Eijnden, E. (2005).
\newblock {Low-order stochastic mode reduction for a realistic barotropic model
  climate}.
\newblock {\em J. Atmos. Sci}, (62):1722--1745.

\bibitem[Franzke et~al., 2015]{Franzke2015}
Franzke, C. L.~E., O'Kane, T.~J., Berner, J., Williams, P.~D., and Lucarini, V.
  (2015).
\newblock {Stochastic climate theory and modeling}.
\newblock {\em Wiley Interdisciplinary Reviews: Climate Change}, 6(1):63--78.

\bibitem[Gallavotti, 2014]{Gallavotti2014a}
Gallavotti, G. (2014).
\newblock {\em {Nonequilibrium and Irreversibility.}}
\newblock Springer International Publishing.

\bibitem[Gallavotti and Lucarini, 2014]{Gallavotti2014}
Gallavotti, G. and Lucarini, V. (2014).
\newblock {\em {Equivalence of Non-equilibrium Ensembles and Representation of
  Friction in Turbulent Flows : The Lorenz 96 Model}}.

\bibitem[Ghil and Childress, 1987]{Ghil1987}
Ghil, M. and Childress, S. (1987).
\newblock {\em {Topics in Geophysical Fluid Dynamics: Atmospheric Dynamics,
  Dynamo Theory, and Climate Dynamics}}, volume~60 of {\em {Applied
  Mathematical Sciences}}.
\newblock Springer New York, New York, NY.

\bibitem[Hallerberg et~al., 2010]{Hallerberg2010}
Hallerberg, S., Paz\'{o}, D., L\'{o}pez, J.~M., and Rodr\'{\i}guez, M.~a.
  (2010).
\newblock {Logarithmic bred vectors in spatiotemporal chaos: Structure and
  growth}.
\newblock {\em Physical Review E - Statistical, Nonlinear, and Soft Matter
  Physics}, 81(6):1--8.

\bibitem[Holton, 2004]{Holton2004}
Holton, J.~R. (2004).
\newblock {\em {An introduction to dynamic meteorology.}}, volume~88.
\newblock Elsevier Academic Press.

\bibitem[Imkeller and von Storch, 2001]{Imkeller2001}
Imkeller, P. and von Storch, J.-S. (2001).
\newblock {\em {Stochastic Climate Models}}, volume~66.
\newblock Birkh\"{a}user Basel, Basel.

\bibitem[Kondrashov et~al., 2017]{Kondrashov2017}
Kondrashov, D., Chekroun, M.~D., and Ghil, M. (2017).
\newblock {Data-adaptive harmonic decomposition and stochastic modeling of
  Arctic sea ice.}
\newblock In {\em {Advances in Nonlinear Geosciences}}. Springer International
  Publishing.

\bibitem[Kravtsov et~al., 2005]{Kravtsov2005}
Kravtsov, S., Kondrashov, D., and Ghil, M. (2005).
\newblock {Multilevel Regression Modeling of Nonlinear Processes: Derivation
  and Applications to Climatic Variability }.
\newblock {\em J. Climate}, 18:4404--4424.

\bibitem[Li et~al., 2012]{Li2012}
Li, F., Rosa, D., Collins, W.~D., and Wehner, M.~F. (2012).
\newblock {Super-parameterization: A better way to simulate regional extreme
  precipitation}.
\newblock {\em Journal of Advances in Modeling Earth Systems}, 4(4):1--10.

\bibitem[Lorenz, 1996]{Lorenz1996}
Lorenz, E.~N. (1996).
\newblock {Predictability - a problem partly solved}.
\newblock In Palmer, T. and Hagedorn, R., editors, {\em {Predictability of
  Weather and Climate}}, pages 40--58. Cambridge University Press.

\bibitem[Lucarini, 2012]{Lucarini2012}
Lucarini, V. (2012).
\newblock {Stochastic Perturbations to Dynamical Systems: A Response Theory
  Approach}.
\newblock {\em Journal of Statistical Physics}, 146(4):774--786.

\bibitem[Lucarini et~al., 2014]{Lucarini2014}
Lucarini, V., Blender, R., Herbert, C., Ragone, F., and Pascale, S. (2014).
\newblock {Mathematical and physical ideas for climate science}.
\newblock {\em Reviews of Geophysics}, 52:809--859.

\bibitem[Lucarini and Sarno, 2011]{Lucarini2011}
Lucarini, V. and Sarno, S. (2011).
\newblock {A statistical mechanical approach for the computation of the
  climatic response to general forcings}.
\newblock {\em Nonlinear Processes in Geophysics}, 18(1):7--28.

\bibitem[Lucarini and Wouters, 2017]{Lucarini2017}
Lucarini, V. and Wouters, J. (2017).
\newblock {Response formulae for n-point correlations in statistical mechanical
  systems and application to a problem of coarse graining}.
\newblock {\em J. Phys. A: Math. Theor.}, 50.

\bibitem[Majda, 2007]{Majda2007}
Majda, A.~J. (2007).
\newblock {Multiscale Models with Moisture and Systematic Strategies for
  Superparameterization}.
\newblock {\em Journal of the Atmospheric Sciences}, 64(7):2726--2734.

\bibitem[Majda et~al., 1999]{Majda1999}
Majda, A.~J., Timofeyev, I., and Vanden-Eijnden, E. (1999).
\newblock {Models for stochastic climate prediction}.
\newblock {\em Proc. Natl. Acad. Sci. USA}, 96:14687--14691.

\bibitem[Majda et~al., 2001]{Majda2001}
Majda, A.~J., Timofeyev, I., and Vanden-Eijnden, E. (2001).
\newblock {A Mathematical Framework for Stochastic Climate Models}.
\newblock {\em Commun. Pure Appl. Math.}, (54):891--974.

\bibitem[Majda et~al., 2003]{Majda2003}
Majda, A.~J., Timofeyev, I., and Vanden-Eijnden, E. (2003).
\newblock {Systematic strategies for stochastic mode reduction in climate}.
\newblock {\em J. Atmos. Sci.}, 60:1705--1722.

\bibitem[McGuffie and Henderson-Sellers, 2005]{McGuffie2005}
McGuffie, K. and Henderson-Sellers, A. (2005).
\newblock {\em {A Climate Modelling Primer}}.
\newblock John Wiley \& Sons, Ltd, Chichester, UK.

\bibitem[Mori et~al., 1974]{Mori1974}
Mori, H., Fujisaka, H., and Shigematsu, H. (1974).
\newblock {A New Expansion of the Master Equation}.
\newblock {\em Progress of Theoretical Physics}, 51(1):109--122.

\bibitem[Neumaier and Schneider, 2001]{Neumaier2001}
Neumaier, A. and Schneider, T. (2001).
\newblock {Estimation of parameters and eigenmodes of multivariate
  autoregressive models}.
\newblock {\em ACM Transactions on Mathematical Software}, 27(1):27--57.

\bibitem[Orrell, 2003]{Orrell2003}
Orrell, D. (2003).
\newblock {Model Error and Predictability over Different Timescales in the
  Lorenz '96 Systems}.
\newblock {\em Journal of the Atmospheric Sciences}, 60(17):2219--2228.

\bibitem[Palmer and Hagedorn, 2006]{Palmer2006}
Palmer, T. and Hagedorn, R. (2006).
\newblock {\em {The predictability of weather and climate}}.
\newblock Cambridge University Press.

\bibitem[Palmer and Williams, 2008]{Palmer2008}
Palmer, T.~N. and Williams, P.~D. (2008).
\newblock {Introduction. Stochastic physics and climate modelling.}
\newblock {\em Philosophical transactions. Series A, Mathematical, physical,
  and engineering sciences}, 366(1875):2421--7.

\bibitem[Park, 2014]{Park2014}
Park, S. (2014).
\newblock {A Unified Convection Scheme ( UNICON ). Part I : Formulation}.
\newblock {\em J. Atmos. Sci.}, 71(Lcl):3902--3930.

\bibitem[Pavliotis and Stuart, 2008]{Pavliotis2008}
Pavliotis, G.~A. and Stuart, A.~M. (2008).
\newblock {\em {Multiscale methods: averaging and homogenization}}.
\newblock Texts in applied mathematics : TAM, Springer, New York, NY.

\bibitem[Peixoto and Oort, 1993]{Peixoto1993}
Peixoto, J. and Oort, A. (1993).
\newblock {\em {Physics of Climate}}.
\newblock American Institute of Physics, New York, NY.

\bibitem[Plant and Yano, 2016]{Plant2016}
Plant, R.~S. and Yano, J.-I. (2016).
\newblock {\em {Parameterization of Atmospheric Convection - Volume 1:
  Theoretical Background and Formulation}}.
\newblock Imperial College Press.

\bibitem[Pope, 2004]{Pope2004}
Pope, S.~B. (2004).
\newblock {Ten questions concerning the large-eddy simulation of turbulent
  flows}.
\newblock {\em New Journal of Physics}, 6.

\bibitem[Ruelle, 1998]{Ruelle1998}
Ruelle, D. (1998).
\newblock {General linear response formula in statistical mechanics, and the
  fluctuation-dissipation theorem far from equilibrium}.
\newblock {\em Physics Letters A}, 245(3-4):220--224.

\bibitem[Ruelle, 2009]{Ruelle2009}
Ruelle, D. (2009).
\newblock {A review of linear response theory for general differentiable
  dynamical systems}.
\newblock {\em Nonlinearity}, 22:855--870.

\bibitem[Sakradzija et~al., 2016]{Sakradzija2016}
Sakradzija, M., Seifert, A., and Dipankar, A. (2016).
\newblock {A stochastic scale-aware parameterization of shallow cumulus
  convection across the convective gray zone}.
\newblock {\em Journal of Advances in Modeling Earth Systems}, 8:786--812.

\bibitem[Schneider and Neumaier, 2001]{Schneider2001}
Schneider, T. and Neumaier, A. (2001).
\newblock {Algorithm 808: ARfit---a matlab package for the estimation of
  parameters and eigenmodes of multivariate autoregressive models}.
\newblock {\em ACM Transactions on Mathematical Software}, 27(1):58--65.

\bibitem[Trevisan et~al., 2010]{Trevisan2010}
Trevisan, A., Isidoro, D., and Talagrand, O. (2010).
\newblock {Four-dimensional variational assimilation in the unstable subspace
  and the optimal subspace dimension}.
\newblock {\em Q. J. R. Meteorol. Soc. 136(647)}, (January):487--496.

\bibitem[Trevisan and Uboldi, 2004]{Trevisan2004}
Trevisan, A. and Uboldi, F. (2004).
\newblock {Assimilation of Standard and Targeted Observations within the
  Unstable Subspace of the Observation -- Analysis -- Forecast Cycle System}.
\newblock {\em J. Atmos. Sci. 61(1)}, pages 103--113.

\bibitem[Wilks, 2005]{Wilks2005}
Wilks, D.~S. (2005).
\newblock {Effects of stochastic parametrizations in the Lorenz '96 system}.
\newblock {\em Quarterly Journal of the Royal Meteorological Society},
  131(606):389--407.

\bibitem[Wouters et~al., 2016]{Wouters2016a}
Wouters, J., Dolaptchiev, S.~I., Lucarini, V., and Achatz, U. (2016).
\newblock {Parametrization of stochastic multiscale triads}.
\newblock {\em Nonlin. Processes Geophys.}, 23:435--445.

\bibitem[Wouters and Gottwald, 2017]{Wouters2017}
Wouters, J. and Gottwald, G. (2017).
\newblock {Edgeworth expansions for slow-fast systems and their application to
  model reduction for finite time scale separation,}.
\newblock {\em arXiv:1708.06984}.

\bibitem[Wouters and Lucarini, 2012]{Wouters2012}
Wouters, J. and Lucarini, V. (2012).
\newblock {Disentangling multi-level systems: averaging, correlations and
  memory}.
\newblock {\em Journal of Statistical Mechanics: Theory and Experiment},
  2012(03):P03003.

\bibitem[Wouters and Lucarini, 2013]{Wouters2013}
Wouters, J. and Lucarini, V. (2013).
\newblock {Multi-level Dynamical Systems: Connecting the Ruelle Response Theory
  and the Mori-Zwanzig Approach}.
\newblock {\em Journal of Statistical Physics}, 151(5):850--860.

\bibitem[Wouters and Lucarini, 2016]{Wouters2016}
Wouters, J. and Lucarini, V. (2016).
\newblock {Parametrization of Cross-scale Interaction in Multiscale Systems}.
\newblock In Chang, C.-P., Ghil, M., Latif, M., and Wallace, J.~M., editors,
  {\em {Climate Change: Multidecadal and Beyond}}, volume~15, pages 67--80.

\bibitem[Young, 2002]{Young2002}
Young, L.~S. (2002).
\newblock {What are SRB measures, and which dynamical systems have them?}
\newblock {\em Journal of Statistical Physics}, 108(5-6):733--754.

\bibitem[Zwanzig, 1960]{Zwanzig1960}
Zwanzig, R. (1960).
\newblock {Ensemble Method in the Theory of Irreversibility}.
\newblock {\em The Journal of Chemical Physics}, 33(5):1338--1341.

\bibitem[Zwanzig, 1961]{Zwanzig1961}
Zwanzig, R. (1961).
\newblock {Memory effects in irreversible thermodynamics}.
\newblock {\em Physical Review}, 124(4):983.

\end{thebibliography}
\end{document}